\documentclass[11pt]{article}

\usepackage{amsmath}
\usepackage{amssymb}
\usepackage{hyperref}
\usepackage[usenames, dvipsnames]{color}
\usepackage{mathrsfs}

\hypersetup{
    unicode=true,           % non-Latin characters in Acrobat??s bookmarks
    pdftoolbar=false,       % show Acrobat??s toolbar?
    pdfmenubar=false,       % show Acrobat??s menu?
    pdffitwindow=false,     % window fit to page when opened
    pdfstartview={FitH},    % fits the width of the page to the window
    pdftitle={},            % title  
    pdfauthor={},           % author
    pdfsubject={},          % subject of the document
    pdfcreator={},          % creator of the document
    pdfproducer={},         % producer of the document
    pdfkeywords={},         % list of
    pdfnewwindow=true,      % links in new window
    colorlinks=true,        % false: boxed links; true: colored links
    linkcolor=black,        % color of internal links
    citecolor=blue,         % color of links to bibliography
    filecolor=green,        % color of file links
    urlcolor=blue           % color of external links
}

.tfm scaled 1000 %  BGotyk
.tfm scaled 1000 %  Gotyk
.tfm scaled 1000 %  Kaligrafik

\newtheorem{Prop}{Proposition}[section]

\newcommand{\To}{\longrightarrow}

\def \a {\alpha}
\def \b {\beta}

\def \d {\delta}

\def \g {\gamma}
\def \et {\eta}

\def \dl{\Delta}

\def \dd{\mathrm{d}}

\newcommand{\ex}[1]{\mathrm{e}^{#1}}

\newcommand \bra[1]{\langle {#1} |}
\newcommand \ket[1]{|{#1} \rangle}

\newcommand{\pbinom}[2]{\cbr{\substack{\scriptscriptstyle{#1}\\\scriptscriptstyle{#2}} }}

\newcommand{\vertex}[4]{\gamma_{#1}\left[\substack{#2\\#3}\right]_{#4}}
\newcommand{\bform}[4]{\beta_{#1}\left[\substack{#2\\#3}\right]^{#4}}

\newcommand{\igram}[4]{\big({G^{(#1)}_{#2}}\big)^{#3\,#4}}
\newcommand{\clfourptblock}[5]{f_{#1}\sbr{\begin{smallmatrix}
                                   #4 & #3
                                   \\
                                   #5 & #2
                                   \end{smallmatrix}}}
\newcommand{\fourptblock}[5]{\mathcal{F}_{#1}\sbr{\begin{smallmatrix}
                                   #4 & #3
                                   \\
                                   #5 & #2
                                   \end{smallmatrix}}}                           
\newcommand{\fourptblcoeff}[6]{\mathcal{F}^{(#1)}_{#2}\sbr{\begin{smallmatrix}
                                   #5 & #4
                                   \\
                                   #6 & #3
                                   \end{smallmatrix}}}
\def\hyper{{{}_{\sf\scriptscriptstyle 2}\mbox{\sf F}_{\!\sf\scriptscriptstyle 1}} }

\newcommand \ord[1]{\mathcal{O}({#1})}
\newcommand \cbr[1]{\left({#1}\right)}
\newcommand \sbr[1]{\left[{#1}\right]}

\begin{document}
\title{\LARGE\sf Solving Heun's equation using conformal blocks}
\author{\sf Marcin Pi\c{a}tek\footnote{e-mail: piatek@fermi.fiz.univ.szczecin.pl, 
radekpiatek@gmail.com}$^{\;\;\;a,\,c}$
\hspace{1cm}
Artur R.~Pietrykowski\footnote{e-mail: pietrie@theor.jinr.ru}$^{\;\;\;b,\,c}$
\\[8pt]
${}^a$ Institute of Physics, University of Szczecin\\
Wielkopolska 15, 70--451 Szczecin, Poland
\\[8pt]
${}^b$ Institute of Theoretical Physics, University of Wroc{\l}aw\\ 
pl.~M.~Borna, 950-204 Wroc{\l}aw, Poland
\\[8pt]
${}^c$ Bogoliubov Laboratory of Theoretical Physics,\\
Joint Institute for Nuclear Research, 141980 Dubna, Russia}
\date{}
\maketitle

\begin{abstract}\noindent
It is known that the classical limit of the second order BPZ null vector decoupling equation for the 
simplest two 5-point degenerate spherical conformal blocks yields: 
(i) the normal form of the Heun equation with the 
complex accessory parameter determined by the 4-point classical block on the sphere, and (ii) a pair of 
the Floquet type linearly independent solutions.
A key point in a derivation of the above result is the classical asymptotic of 
the 5-point degenerate blocks in which the so-called heavy and light contributions decouple.
In the present work the semi-classical heavy-light factorization of the 5-point degenerate conformal 
blocks is studied. In particular, a mechanism responsible for the decoupling of the heavy and light 
contributions is identified. Moreover, it is shown that the factorization property yields a practical method of 
computation of the Floquet type Heun's solutions. 
Finally, it should be stressed that tools analyzed in this work have a broad spectrum of applications, in 
particular, in the studies of spectral problems with the Heun class of potentials, sphere-torus correspondence 
in 2d CFT, the KdV theory, the connection problem for the Heun equation and black hole physics. 
These applications are main motivations for the present work.
\end{abstract}

\newpage\tableofcontents

\section{Introduction}
\subsection{The Heun equation}
To begin with let us consider the Heun equation, i.e., the Fuchsian equation with
four singularities. The most familiar form of the Heun equation reads as follows
\cite{RA}\footnote{Eq.~(\ref{mff}) is called
now the Heun equation in honor of its first investigator --- German mathematician Karl Heun (1859--1929).}
\begin{equation}\label{mff}
\frac{{\rm d}^2 \Phi}{{\rm d}z^2}+
\left[\frac{\gamma}{z}+\frac{\omega}{z-1}+\frac{\epsilon}{z-x}\right]\frac{{\rm d} \Phi}{{\rm d}z}
+\frac{\alpha\beta z-{\sf Q}}{z(z-1)(z-x)}\Phi=0.
\end{equation}

Eq.~(\ref{mff}) has four regular singular points at $0,1,x$ and $\infty$. 
In (\ref{mff}) it is assumed the condition $\epsilon=\alpha+\beta-\gamma-\omega+1$ needed
to ensure regularity of the point at $\infty$. The complex number ${\sf Q}$ is called the accessory 
parameter. Another representation of the Heun equation known as the {\it normal form} looks as follows 
\cite{DLMF}
\begin{eqnarray}\label{F3}
\frac{{\rm d}^{2}\Psi}{{\rm d}z^2}-\left[\frac{\rm A}{z}+\frac{\rm B}{z-1}+\frac{\rm C}{z-x}
+\frac{\rm D}{z^2}+\frac{\rm E}{(z-1)^2}+\frac{\rm F}{(z-x)^2}\right]\Psi = 0.
\end{eqnarray}
Eqs.~(\ref{mff}) and (\ref{F3}) are linked by the substitution
$\Psi(z)=z^{\gamma/2}(z-1)^{\omega/2}(z-x)^{\epsilon/2}\,\Phi(z)$,
where
$$
{\rm A} = -\frac{\gamma\omega}{2}-\frac{\gamma\epsilon}{2x}+\frac{{\sf Q}}{x},
\;\;\;\;\;\;\;
{\rm B} = \frac{\gamma\omega}{2}-\frac{\omega\epsilon}{2(x-1)}-\frac{{\sf Q}-\alpha\beta}{x-1},
\;\;\;\;\;\;\;
{\rm C} = \frac{\gamma\epsilon}{2x}+\frac{\omega\epsilon}{2(x-1)}-\frac{x\alpha\beta-{\sf Q}}{x(x-1)}
$$
and
\begin{eqnarray*}\label{p12}
{\rm D}=\tfrac{1}{2}\gamma\left(\tfrac{1}{2}\gamma-1\right),\;\;\;\;\;\;\;\;\;\;
{\rm E}=\tfrac{1}{2}\omega\left(\tfrac{1}{2}\omega-1\right),\;\;\;\;\;\;\;\;\;\;
{\rm F}=\tfrac{1}{2}\epsilon\left(\tfrac{1}{2}\epsilon-1\right).
\end{eqnarray*}

In the present paper we study 
a realization of the Heun equation in a two-dimensional conformal field theory 
($\rm CFT_{2}$) and discuss some of its consequences. 
In $\rm CFT_{2}$, the Heun equation occurs in the normal form (\ref{F3}) usually written as
\begin{equation}
\label{Fuchss2bis}
\frac{{\rm d}^{2}\Psi}{{\rm d}z^2}+\left[\frac{\delta_1}{z^2}
+\frac{\delta_2}{(z-x)^2} +\frac{\delta_3}{(1-z)^2}
+\frac{\delta_1+\delta_2+\delta_3-\delta_4}{z(1-z)}
+\frac{x(1-x)c_{2}(x)}{z(z-x)(1-z)}\right]\Psi=0.
\end{equation}
From (\ref{F3}) and (\ref{Fuchss2bis}) one has
\begin{eqnarray}%\label{NormalForm}
-{\rm A} &\!=\!& \delta_1+\delta_2+\delta_3-\delta_4+c_2(x) (x-1),
\;\;\;\;\;\;\;\;
-{\rm B} \;=\; \delta_4-\delta_1-\delta_2-\delta_3-c_2(x)x,\nonumber\\
-{\rm C}&\!=\!& c_2(x),\;\;\;\;\;-{\rm D}=\delta_1,\;\;\;\;\;-{\rm E}=\delta_3,\;\;\;\;\;-{\rm F}=\delta_2.
\end{eqnarray}
Here we just only briefly announce that
eq.~(\ref{Fuchss2bis}) emerges in {\it Liouville field theory} (LFT) and 
in a model independent {\it chiral} $\rm CFT_{2}$. These two 
contexts are related to computations of certain {\it conformal blocks}.
For this reason, before we express our concrete 
goals and motivations of the present work, we would like to spell out basic 
information about the {\it quantum} and {\it classical}  
conformal blocks. In particular, we will list some interesting current research topics related to them. 

\subsection{Quantum and classical conformal blocks}
Let $C_{g,n}$ be the Riemann surface with genus $g$ and $n$ punctures.
The basic objects of any two-dimensional conformal field theory 
living on $C_{g}$ \cite{Belavin:1984vu,Eguchi:1986sb}
are the $n$-point correlation functions of primary physical vertex operators
defined on $C_{g,n}$. Any correlation function can
be factorized according to the pattern given by the pant decomposition of
$C_{g,n}$ and written as a sum  (or an integral for theories with a continuous spectrum)
which includes the terms consisting of the holomorphic and anti-holomorphic
conformal blocks times the 3-point functions of the model
for each pair of pants. The Virasoro conformal block
${\cal F}_{c,\Delta_p}[\Delta_i]({\sf Z})$ on $C_{g,n}$
depends on the cross ratios of the vertex operators locations denoted symbolically
by ${\sf Z}$ and on the $3g-3+n$ intermediate conformal weights 
$\Delta_p$. Moreover, it depends on the $n$ external conformal weights
$\Delta_i$ and on the central charge $c$.

In the operator formalism \cite{Moore:1988uz,Felder1990}
the conformal blocks on the Riemann sphere are defined as matrix elements  
(vacuum expectation values)
of radially ordered compositions of the primary chiral vertex operators (CVO's)
acting between Verma modules --- the highest weight representations of the Virasoro algebra. 
Conformal blocks on the torus are traced cylinder matrix elements of
2d Euclidean `space-time' translation operators and CVO's insertions, i.e., 
`chiral partition functions'.
Conformal blocks are fully determined by the underlying conformal symmetry.
These functions possess an interesting, although not yet completely understood analytic
structure. In general, they can be expressed only as a formal power series and
no closed formula is known for its coefficients.

Among  the issues concerning
conformal blocks which are still not fully understood there is the problem
of their {\it classical limit} \cite{Zamolodchikov:1995aa}.
This is the limit in which all parameters of the conformal blocks
tend to infinity in such a way that their ratios are fixed:
$$
\Delta_i,\; \Delta_p,\; c\;\To\;\infty,
\;\;\;\;\;\;\;\;\;\;\;\;
\frac{\Delta_i}{c}\;=\;\frac{\Delta_p}{c}\;=\;{\rm const.}\;\;\;.
$$
For the standard parametrization of the central charge $c=1+6Q^2$, 
where $Q=b+\frac{1}{b}$ and for `heavy' weights 
$(\Delta_p,\Delta_i)=\frac{1}{b^2}(\delta_p,\delta_i)$ with $\delta_p,\delta_i={\cal O}(b^0)$
the classical limit corresponds to $b\to 0$.
There exist many convincing arguments 
that in the classical limit the conformal blocks behave exponentially with respect 
to ${\sf Z}$:
\begin{equation*}
{\cal F}_{c,\Delta_p}[\Delta_i]({\sf Z})\;\stackrel{b\to 0}{\sim}
\;{\rm e}^{\frac{1}{b^2}f_{\delta_p}\left[\delta_i\right]({\sf Z})}.
\end{equation*}
The functions $f_{\delta_p}\left[\delta_i\right]\!({\sf Z})$ are known as 
the classical conformal blocks \cite{Zamolodchikov:1995aa,Hadasz:2005gk}.

Conformal blocks may also include the `light' conformal weights 
${\Delta}_{\rm light}$ which appear as contributions from the `light' 
chiral vertex operators 
$V_{\Delta_{\rm light}}(y)$. The light conformal weights are defined by the property 
$\lim_{b\to 0}b^2{\Delta}_{\rm light}=0$.
It is known (but not proven in general) that light insertions have no influence 
to the classical limit,
i.e.,~do not contribute to the classical blocks:  
\begin{equation}\label{HL}
\left\langle\prod_{i=1}^{n}V_{\Delta_{\rm heavy}}(z_i) V_{\Delta_{\rm light}}(y)\right\rangle
\;\stackrel{b\to 0}{\sim}\;\Psi(y)\;{\rm e}^{\frac{1}{b^2}f_{\delta_p}\left[\delta_i\right]({\sf Z})}.
\end{equation}

Recently, a considerable progress in the theory of conformal blocks and their
applications has been achieved. This is mainly due to the discovery of the so-called AGT correspondence 
\cite{Alday:2009aq}. 
The AGT conjecture states that the Liouville field theory 
correlators on the Riemann surface $C_{g,n}$ with 
genus $g$ and $n$ punctures can be identified with the
partition functions of a class 
$T_{g,n}$ of four-dimensional ${\cal N}=2$ 
supersymmetric SU(2) quiver gauge
theories. A significant part of the AGT conjecture 
is an exact correspondence between the Virasoro
blocks on $C_{g,n}$ and the instanton sectors of the Nekrasov 
partition functions of the gauge theories
$T_{g,n}$. Soon after its discovery, the AGT hypothesis has been
extended, in particular, (i)~to the correspondence between conformal
Toda correlators and SU(N) gauge
theories partition functions \cite{Wyllard:2009hg}; (ii)~to the relation (cf.~\cite{Gaiotto:2009}) between 
{\it irregular conformal blocks} 
\cite{Marshakov:2009,Alba:2009fp,Gaiotto:2012sf} and Nekrasov's instanton partition functions
for `non-conformal' ${\cal N}=2$, SU(2) super Yang--Mills
theories.

The irregular conformal blocks discovered by Gaiotto were introduced in his work~\cite{Gaiotto:2009}
as products of some new states belonging to the Hilbert space of $\rm CFT_{2}$.
These novel irregular Gaiotto states are kind of coherent vectors for some Virasoro 
generators. It is also known that irregular blocks 
can be obtained form standard (regular) conformal blocks in
properly defined {\it decoupling limits} of the external conformal weights, 
cf.~\cite{Marshakov:2009,Alba:2009fp}. Furthermore, the Gaiotto vectors can be
understood as a result of suitable defined {\it collision limit} of locations of vertex 
operators in their operator product expansion (OPE), cf.~\cite{Gaiotto:2012sf}.

Interestingly, the classical limit exists also for irregular blocks and 
consistently defines the {\it classical irregular blocks}. 
This claim first time has appeared in \cite{Piatek:2014lma} as a result of 
non-conformal AGT relations and yet another new duality, the so-called Bethe/gauge correspondence 
\cite{NS:2009}.

Let us recall that the AGT correspondence works at the level of the quantum Liouville
field theory. At this point it arises the question as to what happens if we 
proceed to the classical limit of the Liouville theory. It turns out that the 
semi-classical limit of the LFT correlation functions 
\cite{Zamolodchikov:1995aa} corresponds to the
Nekrasov--Shatashvili (NS) limit of the Nekrasov partition functions \cite{NS:2009}.
In particular, a consequence of that correspondence is that the classical
conformal blocks can be identified with the instanton sectors of the
{\it effective twisted superpotentials} 
\cite{Nekrasov:2011bc,Teschner:2010je}.\footnote{On the relation
between the quantisation of the Hitchin system, SYM theories, $\rm CFT_{2}$ and the
geometric Langlands program, see also \cite{Teschner:2017djr}.} 
The latter quantities determine the low energy effective
dynamics of the two-dimensional gauge theories restricted to the
so-called $\Omega$-background. The twisted superpotentials
play also a pivotal role in another duality, the aforementioned 
Bethe/gauge  correspondence \cite{NS:2009}
that maps supersymmetric vacua of the ${\cal N}=2$ theories to Bethe
states of quantum integrable systems (QIS). 
A result of that duality is that
the twisted superpotentials are identified with the {\it Yang--Yang functions}
which describe spectra of the corresponding quantum integrable systems.
Hence, combining the classical--NS limit of the AGT duality and the Bethe/gauge
correspondence one thus gets the triple correspondence: 
\begin{center}
\begin{tabular}{|c|c|c|}
\hline\hline
{$\rm CFT_{2}$} & ${\rm 2d}\;{\cal N}=2\;$SU(2)\;SYM & {\rm 2-particle}\;{\rm QIS}\;\\ \hline\hline
classical Virasoro blocks &\;\;\;\;twisted superpotentials  & spectra of Schr\"{o}dinger operators              \\ \hline
\end{tabular}
\end{center}
which links the 
classical Virasoro blocks to the SU(2) 
twisted superpotentials and then to spectra of some Schr\"{o}dinger operators.
Indeed, let us note that 2-particle QIS are nothing but
quantum--mechanical systems. The above correspondence can be extended to the following: 
\begin{center}
\begin{tabular}{|c|c|c|}
\hline\hline
{$\rm CFT_{2}$} & ${\rm 2d}\;{\cal N}=2$\;SU(N)\;{\rm SYM} & {${\rm N}$-particle}\;{\rm QIS}\;\\ \hline\hline
classical Toda blocks &\;\;\;\;  twisted superpotentials  & Yang--Yang functions              \\ \hline
\end{tabular}
\end{center}
if we take the classical--NS limit of the generalized AGT conjecture \cite{Wyllard:2009hg}.

Concluding, the interest in the classical conformal blocks and their
uses has recently dramatically increased. 
Almost every day one may find in a literature more new, 
additional to those described above, fascinating contexts, in which these functions emerge. Some of these 
topics are: spectral problems for some Schr\"{o}dinger operators 
\cite{Piatek:2014lma,Piatek:2015jva,Piatek:2016xhq,Basar:2015xna};
the Painlev\'{e} VI equation \cite{Litvinov:2013sxa}; the KdV equation 
\cite{He:2014yka};
the isomonodromic deformation problem \cite{Teschner:2017rve};
S-duality in ${\cal N}=2$ super Yang--Mills theories 
\cite{Kashani-Poor:2014mua};
entanglement entropy in ${\rm CFT_{2}}$ and ${\rm CFT_{2}/AdS_3}$ holography 
\cite{Hartman:2013mia,Asplund:2014coa};
quantum chaos \cite{Roberts:2014ifa}; the connection problem for the Heun 
equation and 
scattering in black hole backgrounds 
\cite{daCunha:2015fna,daCunha:2015ana,Novaes:2014lha};
black holes, holography and information paradox \cite{Anous:2016kss}; 
semi-classical spectra and restrictions on holographic two-dimensional CFT's 
\cite{Asplund:2014coa,Chang:2015qfa};
holographic interpretation of classical blocks and classical bootstrap
\cite{Chang:2015qfa,Alkalaev:2016rjl,Alkalaev:2016ptm,Alkalaev:2015fbw,
Alkalaev:2015lca,Alkalaev:2015wia,Chang:2016ftb,Banerjee:2016qca}.

\subsection{Objectives, motivations, outline} 
As has been already mentioned, in two-dimensional conformal field theory the Heun equation emerges in two 
circumstances.

First, eq.~(\ref{Fuchss2bis})
one gets in the classical limit from the BPZ null vector decoupling 
equation \cite{Belavin:1984vu} 
for the Liouville 5-point function of primary fields with one light degenerate field 
${\sf V}_{\Delta_{2,1}}(z,\bar z)$, where $\Delta_{2,1}=-\frac{1}{2}-\frac{3}{4}\,b^2$.
The accessory parameter $c_{2}(x)$ in eq.~(\ref{Fuchss2bis}) is determined in this framework 
by the classical Liouville action on the 4-punctured sphere.  

On the other hand, the degenerate 5-point blocks in the Liouville 
correlation function fulfill the same null 
vector decoupling equation as the physical correlator itself.
Applying to this equation the asymptotic (\ref{HL}) with $n=5$ 
and $\Delta_{\rm light}=\Delta_{2,1}$ one gets the normal form Heun
equation with the two linearly independent solutions $(\Psi_+, \Psi_-)$ known as the 
{\it path--multiplicative} or {\it Floquet type solutions}, cf.~\cite{Litvinov:2013sxa,Slavyanov}. 
Here, the Heun accessory parameter $c_{2}(x)$ is a complex-valued function
determined by the classical 4-point block on the sphere. In fact, one obtains the following claim:

\begin{Prop}\label{P1}
%{\it
%{\bf Proposition 1}
{\sf The accessory parameter $c_{2}(x)$ in the Heun equation (\ref{Fuchss2bis}) is given by the derivative
of the classical 4-point block on the Riemann sphere w.r.t. the modular parameter $x$, 
\begin{equation}\label{c2}
c_{2}(x) =  {\partial\over \partial x}\,
f_{\delta}\!\left[_{\delta_{4}\;\delta_{1}}^{\delta_{3}\;\delta_{2}}\right]\!(x),
\;\;\;\;\;\;\;\;
\delta:=\tfrac{1}{4}\left(1-\lambda^2\right)
\end{equation}
iff the two linearly independent solutions $(\Psi_{+}(z), \Psi_{-}(z))$ 
of the equation (\ref{Fuchss2bis}) are of the Floquet type and
have diagonal monodromy along a curve $\gamma(0,x)$ 
encircling both $0$ and $x$, and the corresponding monodromy matrix 
${\mathrm M}_{\gamma(0,x)}$ obeys ${\rm Tr}\,{\mathrm M}_{\gamma(0,x)}=-2\cos(\pi\lambda)$.}
%\hfill\(\bullet\)
%}
\end{Prop}

Recall that the path--multiplicative (Floquet) solutions of the Heun equation are related to two singularities, 
for example, $z=0$ and $z=1$. Let $\gamma$ be a smoothly parametrized 
closed contour encircling these points: $\gamma:z=\varphi(s)$ for $s\in[0,1]$, 
$\varphi(0)=\varphi(1)=z_0$. A Floquet solution $y(z)$ is defined as a solution satisfying 
$y(\varphi(1))={\rm e}^{2\pi i\sigma}y(\varphi(0))$. So, it is a solution which is 
multiplied by a constant factor ${\rm e}^{2\pi i\sigma}$ if we pass around a simple closed contour 
in the $z$-plane that encircles two of the three singularities $0,1,x$.
The value $\sigma$ is called a {\it Floquet exponent} or {\it path exponent} or {\it characteristic exponent}.
It depends on the parameters of the Heun equation \cite{Slavyanov}. Within the CFT
realization the characteristic exponent $\sigma$ is determined by the 
intermediate classical conformal weight $\delta$.

To get the claim above one can use perturbative methods 
(see \cite{Menotti:2014kra,Menotti:2016jut}, where the implication $\Leftarrow$ was shown)
and a standard conformal field theory machinery.
It should be stressed that the CFT tools, such as the OPE, allows to find monodromy properties of the solutions 
$(\Psi_{+}(z), \Psi_{-}(z))$ without their explicit computation.
However, it would be interesting to know if 
these solutions can be calculated within CFT$_2$.
For this reason in the present paper we study a mechanism of the semi-classical 
`heavy--light' factorization (\ref{HL}) in the case of the degenerate 5-point blocks. Our aim here is
to answer the question whether the formula (\ref{HL}) for $n=5$ and $\Delta_{\rm light}=\Delta_{2,1}$ 
really allows to explicitly compute the Floquet type Heun's solutions. 

Indeed, little is known about a concrete form of the Floquet type solutions (cf.~the end of 
subsection 3.2.1 in \cite{Slavyanov}) and an 
elaboration of any method useful for practical numerical calculation is of great 
importance.
In particular, in \cite{Slavyanov} one can read that the path-multiplicative 
Heun's functions `theoretically can be constructed as series' 
\begin{equation}\label{Hp}
\sum_{-\infty}^{+\infty}c_n f_{\sigma+n}(z)
\end{equation}
with conjectured form of the coefficients
$f_{\sigma+n}(z)$ given by the hypergeometric function 
%$f_{\sigma+n}(z)=z^{\sigma}\hyper(\cdot;\cdot;\cdot;z)$,
$f_{\sigma+n}(z)=z^{\sigma}{_{2}{\sf F}}_{1}(\cdot;\cdot;\cdot;z)$,
where dots denote dependence on the
combinations of parameters of the Heun equation and the characteristic exponent $\sigma$.
However, further one can find in \cite{Slavyanov} that `in the general case, the form
of the dependence of the path exponents $\sigma$ on the parameters of 
the Heun equation is lacking, and hence expansions (\ref{Hp})
cannot be used for practical numerical calculations.'

The answer to the above question allows to make contact with the 
perturbative techniques used to solve the 
monodromy problem for the Heun equation determining the classical 4-point block, 
cf.~\cite{Menotti:2014kra,Menotti:2016jut}.
Both, the CFT and complementary perturbative methods can be capitalized in concrete applications.
In particular, it seems very promising to use the CFT methods 
in the study of scalar perturbations of certain black hole (BH) backgrounds described by the Heun 
equation. This particular application is our main motivation for the present work. We spell out it in more 
details below. 

It is known that the Klein--Gordon equation in the Kerr--${\rm AdS_{5}}$ background \cite{HHT}
can be reduced to the two (angular and radial) Heun 
equations by a separation of variables \cite{KL}.\footnote{In fact, a bit more general situation
can be considered, than those in \cite{KL}, namely, for the 
metric of the 5-dimensional {\it charged} AdS BH, cf.~e.g.~\cite{AD}.
Such BH solution contains the electric charge $Q$ in addition and for $Q=0$ reduces to the metric 
discussed in \cite{HHT,KL}.} 
Having in mind two contexts in which the Heun equation appears in 
2-dimensional CFT one may consider an idea of the formal 
correspondence between the dynamics of a scalar field in 
Kerr--${\rm AdS_{5}}$ and the Liouville theory/chiral ${\rm CFT_{2}}$. Such concept has 
recently occurred in \cite{Amado:2017kao}.\footnote{See also  
\cite{daCunha:2015fna,daCunha:2015ana,Novaes:2014lha}.}
First, it would be nice
to explore this relation at the formal level, i.e.~to complete a dictionary 
and to describe formal implications, as it has been partially done in \cite{Amado:2017kao}.
Interestingly, this identification has many common features with what is known as the Kerr/CFT correspondence 
\cite{GHSS}. 
It would be very interesting to examine whether it is an accidental similarity or indeed something deeper.

Then, one may ask about its possible application in the computation of the so-called BH
quasi-normal modes (QNMs), cf.~\cite{BCS}. 
These quantities are certain {\it complex frequency modes} that resonate when the 
BH is weakly perturbed, e.g.~by the weak scalar field. The quasi-normal modes are important for studying 
a stability of the black holes. They are also useful in the study of a plasma through the finite temperature 
AdS/CFT correspondence. For the Kerr--${\rm AdS_{5}}$ BH a 
preliminary analysis shows that the parameter in the radial and angular Heun 
equations, which can be interpreted as the frequency of the BH oscillations, 
is determined by the complex accessory 
parameter, and so, by the {\it calculable} 
complex-valued classical 4-point block 
$f_{\delta}\!\left[_{\delta_{4}\;\delta_{1}}^{\delta_{3}\;\delta_{2}}\right]\!(x)$. 
Using such an identification one can express the frequency as a function of the other 
parameters in the radial/angular Heun equation and the characteristic exponent related to $\delta$. 
However, to interpret the frequency parameter as the BH frequency QNMs the scalar field, and hence, its radial 
and angular parts have to fulfill physically appropriate boundary conditions at the horizon 
(purely ingoing waves) 
and at the spatial 
infinity (purely outgoing waves).\footnote{Exactly, the presence of the horizon implies that the boundary value 
problem, which
must be solved to determine BH QNMs, is non-hermitian and associated eigenvalues (= QNMs) are complex.}
Therefore, one of the key points that needs to be 
clarified here is to answer the question whether the solutions available within CFT$_2$  
obey these boundary conditions.

The above ideas should work also for other BH backgrounds. 
For instance, the radial equation for the scalar perturbation of the Kerr--NUT--(A)dS BH
in four dimensions is reducible to the Heun equation.
Next, {\it confluent} Heun equations determine the scalar perturbations of the Kerr metric in 
four dimensions.\footnote{These equations are known in the literature as the Teukolsky equations 
(cf.~\cite{Slavyanov} and refs.~therein).}
It is an interesting question of how to get the latter, i.e., the 
confluent Heun equation together with its certain 
solutions entirely within the formalism of CFT$_2$. Perhaps one needs to analyze the collision and classical 
limits of the BPZ equation for the 
5-point blocks. In such a way the quantum and classical irregular conformal blocks should come to the game, 
cf.~\cite{daCunha:2015ana}.\footnote{Moreover, it is reasonable to expect that there exists a monodromy 
problem for the confluent Heun equation whose solution matches the CFT results.}

Closely related to the scattering problems in the black hole backgrounds is the 
connection problem for the Heun and confluent Heun equations, cf.~e.g.~\cite{Slavyanov,CLMR}.
It is a question of how to express any of the pairs  of linearly 
independent solutions of eq.~(\ref{mff}) around some singular point $0$ or $1$ or $x$ or $\infty$ 
as a linear combination of any one of the other three such pairs.
Recently some progress in the solution of the connection problem has been achieved mainly due to the application 
of the so-called `isomonodromic approach'
and results relating regular and irregular conformal blocks to Painlev\'{e} VI and Painlev\'{e} V 
$\tau$--functions, cf.~\cite{daCunha:2015fna,daCunha:2015ana,Novaes:2014lha}. 

The structure of the paper is as follows.
In subsection 2.1 a well-known fact is remind, namely, it is shown how the CFT machinery determines a 
monodromy of the Heun's solutions referred to in the Proposition \ref{P1}. 
In subsection 2.2 the Heun equation is studied within purely mathematical representation 
theoretic formalism of the conformal blocks. 
The formula for the linearly independent solutions of the normal form Heun equation in 
terms of the classical limit of the conformal blocks is derived. 
In subsection 3.1 the heavy-light factorization 
property of the degenerate 5-point blocks is studied. 
This analysis leads to a computation of the limit defining 
the Floquet type Heun's functions and in fact yields a new method of the calculation of the latter. 
A concrete example of the path-multiplicative Heun function computed in this way is
presented there and in the appendix 
\ref{App_HeunCoeffs}. In subsection 3.2 the $x\to 0$ limit of the solution extracted
from the conformal blocks is computed. This calculation confirms that the `CFT solution'
of the Heun equation has for $x\to 0$ the expected form in terms of a hypergeometric function.
In subsection 3.3 a comparison with the perturbative techniques is discussed. 
In section 4 conclusions of the present work and open problems for further research are collected.

\section{The Heun equation in CFT$_{\bf 2}$}
\subsection{Classical limit of BPZ equation for the degenerate five--point function}
\label{LFT5pt}
Let us consider the projected 5-point function on the sphere in the diagonal theory 
$(\Delta_i=\bar\Delta_i)$:
$$
G_{\Delta}(z,x) \;:=\; \Big\langle {\sf V}_4(\infty,\infty){\sf V}_3(1,1)
{\sf V}_{-\frac{b}{2}}(z,\bar z){\sf P}_{\Delta,\Delta}
{\sf V}_2(x,\bar x){\sf V}_1(0,0) \Big\rangle\;,
$$
where ${\sf V}_{\alpha=-\frac{b}{2}}$ is the degenerate field with the conformal weight
$$
\Delta_{-\frac{b}{2}}\;=\;\Delta_{\alpha=-\frac{b}{2}}
\;=\;\alpha(Q-\alpha) \;=\;-\frac{1}{2}-\frac{3}{4}\,b^2,
\;\;\;\;\;\;Q\;=\;b+\frac{1}{b}
$$
and ${\sf V}_{i}$'s  are the four heavy primary operators
($\Delta_i\!=\!b^{-2}\,\delta_i$, $\delta_i\!=\!{\cal O}(1)$). 
The function $G_{\Delta}(z,x)$ satisfies the following null vector decoupling 
(NVD) equation \cite{Belavin:1984vu}:
\begin{eqnarray}
\label{Fuchs1} && \left[ \frac{\partial^2}{\partial z^2}
-b^2\left(\frac{1}{z} - \frac{1}{1-z}\right)\frac{\partial}{\partial
z} \right] G_{\Delta}(z,x)
=
\\
\nonumber &&\hspace{-20pt}-b^2 \left[\frac{\Delta_1}{z^2} +
\frac{\Delta_2}{(z-x)^2} +  \frac{\Delta_3}{(1-z)^2} +
\frac{\Delta_1 \!+\! \Delta_2 \!+\!\Delta_3
\!+\!\Delta_{-\frac{b}{2}}\!-\!\Delta_4}{z(1-z)} + \frac{x(1-x)}{z(z-x)(1-z)}
\frac{\partial}{\partial x}\right]G_{\Delta}(z,x).
\end{eqnarray}
Our aim now is to consider the classical limit
of the equation above. 
The key point here is an observation that in the limit $b\to 0$ only the operator with
weight $\Delta_{-\frac{b}{2}}$ remains light
($\Delta_{-\frac{b}{2}} = {\cal O}(1)$) and its presence in the
correlation function has no influence on the classical dynamics.
Then, for $b\to 0$
\begin{equation}
\label{a5}
G_{\Delta}(z,x) \;\sim\;
\Psi(z)\,
{\rm e}^{-{1\over b^2}\left(
S^{\rm cl}(\delta_4,\delta_3,\delta) +
S^{\rm cl}(\delta,\delta_2,\delta_1)
-f_{\delta}\!\left[_{\delta_{4}\;\delta_{1}}^{\delta_{3}\;\delta_{2}}\right](x)
-\bar f_{\delta}\!\left[_{\delta_{4}\;\delta_{1}}^{\delta_{3}\;\delta_{2}}\right](\bar x)\right)}.
\end{equation}
Indeed, assuming that the light field does not contribute to the classical limit 
we are left with the projected 4-point function of the heavy operators:
\begin{eqnarray*}
&&\hspace*{-2.5cm}
\left\langle
{\sf V}_{4}(\infty,\infty){\sf V}_{3}(1,1) {\sf P}_{\Delta,\Delta}
{\sf V}_{2}(x,\bar x){\sf V}_{1}(0, 0)
\right\rangle \;=\;\nonumber
\\ 
&=&
C\!\left(\Delta_{4},\Delta_{3},\Delta\right)
C\!\left(\Delta,\Delta_{2},\Delta_{1}\right)
{\cal F}_{1+6Q^2,\Delta}\!\left[^{\Delta_3\ \Delta_2}_{\Delta_4\
\Delta_1}\right]\!(x) 
\bar{\cal F}_{1+6Q^2,\Delta}\!\left[^{\Delta_3\ \Delta_2}_{\Delta_4\
\Delta_1}\right]\!(\bar x) 
\\
&\stackrel{b\to 0}{\sim}&
{\rm e}^{-{1\over b^2}\left(
S^{\rm cl}(\delta_4,\delta_3,\delta) +
S^{\rm cl}(\delta,\delta_2,\delta_1)
-f_{\delta}\!\left[_{\delta_{4}\;\delta_{1}}^{\delta_{3}\;\delta_{2}}\right](x)
-\bar f_{\delta}\!\left[_{\delta_{4}\;\delta_{1}}^{\delta_{3}\;\delta_{2}}\right](\bar x)\right)}.
\end{eqnarray*}
The quantities $S^{\rm cl}(\delta_3,\delta_2,\delta_1)$, known as the classical 3-point actions,
are the classical limits of the structure constants $C\!\left(\Delta_{i},\Delta_{j},\Delta_{k}\right)$.
The functions
\begin{eqnarray}
{\cal F}_{c,\Delta}\!\left[^{\Delta_3\ \Delta_2}_{\Delta_4\
\Delta_1}\right]\!(x) &=&
\begin{picture}(65,30)(0,5)
\put(0,0){\makebox(15,10)
{$
  \Delta_4\;
$}}
\put(15,5){\line(1,0){10}}
\put(25,0){\makebox(15,10){$
\scriptstyle 1
$}}
\put(40,5){\line(1,0){40}}
\put(80,0){\makebox(15,10)[c]
{$
 \scriptstyle x
$}}
\put(32.5,10){\line(0,1){8}}
\put(25,20){\makebox(15,10){$
  \Delta_3
$}}
\put(55,8){\makebox(15,10){$
 \scriptstyle \Delta
$}}
\put(87,10){\line(0,1){8}}
\put(78,23){\makebox(15,10){$
\Delta_2
$}}
\put(95,5){\line(1,0){10}}
\put(106,0){\makebox(15,10){$
\Delta_1
$}}
\end{picture}
\nonumber\\[5pt]
&=&
x^{\Delta-\dl_{2}-\dl_{1}}\left(1+
\sum_{n>0} x^{n}
{\cal F}_{c,\Delta}^{(n)}\!\left[^{\Delta_3\ \Delta_2}_{\Delta_4\
\Delta_1}\right]\right)
\end{eqnarray}
and
\begin{equation}\label{class4pt}
f_{\delta}\!\left[_{\delta_{4}\;\delta_{1}}^{\delta_{3}\;\delta_{2}}\right]\!(x) =
\lim\limits_{b\to 0}b^2\log{\cal F}_{1+6Q^2,\Delta}\!\left[^{\Delta_3\ \Delta_2}_{\Delta_4\
\Delta_1}\right]\!(x)
\end{equation}
are quantum, and classical 4-point blocks on the sphere respectively.
The substitution of eq.~(\ref{a5}) into the NVD eq.~(\ref{Fuchs1}) leads, within the limit $b\to 0$, to 
the normal form of the Heun equation (\ref{Fuchss2bis})
with the accessory parameter $c_{2}(x)$
determined by the classical 4-point block, according to (\ref{c2}), cf.~\cite{Litvinov:2013sxa}.

Alternatively, the parameter $c_{2}(x)$ 
can be found once the solution of certain Bethe-like saddle point equation defining the twisted superpotential 
of the SU(2) ${\rm N}_f=4$ ${\cal N}=2$ SYM theory is known, cf.~\cite{Ferrari:2012gc}.

Let us stress that the holomorphic accessory parameter 
$c_{2}(x)$ is related to that 
which emerges in the uniformization theory of the 4-punctured sphere.
The real-valued accessory parameter for $C_{0,4}$ can be found by taking
the classical limit of the NVD equations obeyed by the physical Liouville 5-point function on the sphere
with ${\sf V}_{\alpha=-\frac{b}{2}}$, 
cf.~\cite{Zamolodchikov:1995aa,Hadasz:2005gk,Ferrari:2012gc,Hadasz:2006rb}.
The latter is nothing but the  
projected correlation function discussed above integrated over 
the Liouville field theory spectrum $\Delta=\Delta_\alpha$, $\alpha=\frac{1}{2}Q+i\mathbb{R}^{+}$
with the structure constant $C\!\left(\Delta_{3},\Delta_{2},\Delta_{1}\right)$ 
identified as the Liouville 3-point function 
\cite{Zamolodchikov:1995aa,Dorn:1994xn}.

\subsubsection*{Monodromy properties of the solutions}
Let us consider monodromy properties of the independent solutions $\Psi_{\pm}$  of eq.~(\ref{Fuchss2bis}) 
with the accessory parameter given by (\ref{c2}). 
It turns out that the substitution (\ref{a5}) automatically fixes the solutions 
$\Psi_{\pm}$ to be of the Floquet type with the characteristic exponent $\sigma$ determined by the 
intermediate classical conformal weight $\delta=\frac{1}{4}(1-\lambda^2)$ \cite{Litvinov:2013sxa}.
First, note that the functions on both sides of eq.~(\ref{a5}) should have the same monodromy properties
along the contour encircling the points $0$ and $x$. Secondly, from a completeness of the intermediate 
states follows that the monodromy properties w.r.t.~$z$ of the 5-point function $G_{\Delta}(z,x)$
along a curve encircling both $0$ and $x$ are the same as the monodromy of the 4-point correlator:
\begin{equation}\label{corr}
\Big\langle {\sf V}_4(\infty,\infty){\sf V}_3(1,1)
{\sf V}_{\Delta_{-\frac{b}{2}}}(z,\bar z){\sf V}_{\Delta_\alpha}(0,0)\Big\rangle
\end{equation}
for a curve encircling $0$.
The $z$ dependence (i.e.,~an overall prefactor) of the correlator (\ref{corr})
can be read off from (holomorphic part of) the OPE:
\begin{eqnarray*}
{\sf V}_{-\frac{b}{2}}(z){\sf V}_{\alpha}(0) &=&
z^{\Delta_{+}-\Delta_{-\frac{b}{2}}-\Delta_\alpha}
C\!\left(\Delta_{+},\Delta_{-b/2},\Delta_\alpha\right)
\Big[\,{\sf V}_{\alpha_+}(0)+{\rm descendants}\;\Big]
\\
&+&
z^{\Delta_{-}-\Delta_{-\frac{b}{2}}-\Delta_{\alpha}}
C\!\left(\Delta_{-},\Delta_{-b/2},\Delta_\alpha\right)
\Big[\,{\sf V}_{\alpha_-}(0)+{\rm descendants}\;\Big],
\end{eqnarray*}
where $\Delta_{\pm}\equiv\Delta_{\alpha_{\pm}}$, $\alpha_{\pm}=\alpha\pm\frac{b}{2}$,
and explicitly, 
\begin{eqnarray*}
\Delta_{+}&=&\Delta_\alpha-\alpha b +\frac{b^2}{4}+\frac{1}{2},
\\
\Delta_{-}&=&\Delta_\alpha+\alpha b -\frac{3 b^2}{4}-\frac{1}{2}.
\end{eqnarray*}
Assuming that the intermediate weight $\Delta_\alpha$ is heavy, i.e.,
$$
\alpha\;=\;\frac{1}{2b}(1-\lambda)\;\;\;\;\Leftrightarrow\;\;\;\;
\lim_{b\to 0}b^2\Delta_{\alpha}\;=\;\frac{1}{4}\left(1-\lambda^2\right)\;=\;\delta
$$ 
then in the limit $b\to 0$ one gets
$\Delta_{\pm}-\Delta_{-\frac{b}{2}}-\Delta_{\alpha}\to\frac{1}{2}(1\pm\lambda)$.
Therefore, in the space of solutions of eq.~(\ref{Fuchss2bis}) there exist basis solutions 
$\Psi_{\pm}(z)\varpropto z^{\frac{1}{2}(1\pm\lambda)}$ which analytically continued in $z$
along the path encircling the points $0$ and $x$ satisfy the condition:
\begin{equation}\label{mono}
\Psi_{\pm}({\rm e}^{2\pi i}z)=-{\rm e}^{\pm i\pi\lambda}\Psi_{\pm}(z).
\end{equation}
This corresponds to the monodromy matrix with trace equal $-2\cos(\pi\lambda)$.

As a final remark in this subsection let us notice that knowing monodromies (\ref{mono})
of the solutions of eq.~(\ref{Fuchss2bis}) one can determine the classical 4-point block
solving the Riemann--Hilbert problem formulated as follows: adjust $c_2(x)$
in such a way that the eq.~(\ref{Fuchss2bis}) admits solutions with the monodromy around $0$
and $x$ given by (\ref{mono}). The latter was Zamolodchikov's idea which allowed him to find (i) a large 
classical intermediate weight behavior of the classical 4-point block, (ii) a large quantum intermediate weight 
behavior of the quantum 4-point block and its nome $q(x)$ expansion \cite{Z1}. 

\subsection{Classical limit of BPZ equation for the degenerate five--point blocks}
Let
${\cal V}_{c,\Delta}^{(n)}$ denotes the vector space generated by all vectors of
the form:
\begin{equation}\label{basis}
|\,\Delta^{n}_{I}\,\rangle=L_{-I}|\,{\Delta}\,\rangle 
\equiv L_{-k_{1}}\ldots L_{-k_{\ell(I)}}|\,{\Delta}\,\rangle,
\;\;\;\;\;\;\;\;
n=k_1+\ldots+k_{\ell(I)}=:|I|,
\end{equation}
where $I=(k_{1}\geq\ldots\geq k_{\ell(I)}\geq 1)$ is a partition of $n$,\footnote{We 
will use the notation $I\vdash n$.}
$L_n$'s are the Virasoro generators obeying
\begin{equation}\label{Vir}
[L_n, L_m] = (n-m)L_{n+m}+\frac{c}{12}(n^3 - n)\delta_{n+m,0},
\end{equation}
and $|\,\Delta\,\rangle$ is the highest weight state with the following property:
\begin{equation}\label{hw}
L_0 |\,\Delta\,\rangle = \Delta |\,\Delta\,\rangle,
\;\;\;\;\;\;\;\;\;\;\;
L_n |\,\Delta\,\rangle = 0, \;\;\;\;\forall\;n>0.
\end{equation}
The representation of the Virasoro algebra on the space:
$$
{\cal V}_{c,\Delta}=\bigoplus\limits_{n=0}^{\infty}{\cal V}_{c,\Delta}^{(n)},
\;\;\;\;\;\;\;
{\cal V}_{c,\Delta}^{(0)} = \mathbb{R} |\,\Delta\,\rangle 
$$
defined by the relations (\ref{Vir}), (\ref{hw}) is called the Verma module
with the central charge $c$ and the highest weight $\Delta$. It is clear that
$\dim{\cal V}_{c,\Delta}^{(n)} = p(n)$, where $p(n)$ is the number of partitions of 
$n$ (with the convention $p(0)=1$).
On ${\cal V}_{c,\Delta}^{(n)}$ exists symmetric
bilinear form $\langle\,\cdot\,|\,\cdot\,\rangle$
uniquely defined by the relations
$\langle\,\Delta\,|\,\Delta\,\rangle=1$
and $(L_n)^{\dagger}\;=\;L_{-n}$.

Let $|\,0\,\rangle$ denotes the vacuum state, i.e.,~the highest 
weight state in the vacuum module with the highest weight $\Delta=0$.
Conformal blocks on the Riemann sphere are defined as the matrix elements  
$$
\langle\,0\,|V_{\Delta_n}(z_n)\ldots V_{\Delta_1}(z_1)|\,0\,\rangle
$$
of compositions of the primary chiral vertex operators (CVO's) 
\cite{Moore:1988uz,Felder1990}: 
\begin{eqnarray}\label{cr}
V_{\Delta_j}(z) &\equiv& V_{\alpha_k,\alpha_i}^{\;\alpha_j}(z):
{\cal V}_{\Delta_i}\To {\cal V}_{\Delta_k},
\;\;\;\;\;\;\;\;
\Delta_{l}=\Delta_{\alpha_l}=\alpha_l(Q-\alpha_l), \nonumber
\\
\left[L_n , V_{\Delta}(z)\right] &=& z^{n}\left(z
\frac{\rm d}{{\rm d}z} + (n+1)\Delta
\right)V_{\Delta}(z),\;\;\;\;\;\;\;\;n\in\mathbb{Z}.
\end{eqnarray}
acting between the Verma modules. 

Curious class of CVO's form the degenerate 
operators \cite{Belavin:1984vu} $V_{\Delta_{rs}}(z)$ having conformal weights:
$$
\Delta_{r,s} = \frac{Q^2}{4} - \frac{1}{4}\left(rb + \frac{s}{b} \right)^{2},
\quad\quad Q=b+\frac{1}{b},
\quad\quad r,s\in\mathbb{N}
$$
being zeros of the Kac determinant 
\cite{Kac:1978ge,Feigin:1981st,FeiginFuchs,Thorn:1984sn,Kac:1987gg}:
$$
\det\Big[G_{c,\Delta}^{n}\Big]_{IJ}\;=\;
\det\,\langle\,\Delta_{I}^{n}\,|\,\Delta_{J}^{n}\,\rangle\;=\;
{\rm const.}(n)\times
\prod_{1\leq rs \leq n} (\Delta - \Delta_{r,s})^{p(n-rs)}.
$$
Let us consider the degenerate chiral vertex operators of the form 
\begin{eqnarray}
V^{(\pm)}_{\Delta_{2,1}}(z)\;\equiv\;V^{-b/2}_{\beta_{\pm},\beta}(z), 
&& \Delta_{2,1}=\Delta_{-b/2}=-\frac{1}{2}-\frac{3}{4}b^2.
\end{eqnarray}
If $\beta_{\pm}=\beta\pm \frac{b}{2}$ then

\smallskip\noindent
1.~the CVO's $V^{(\pm)}_{\Delta_{2,1}}(z)$ satisfy the differential equation
\begin{equation}
\frac{1}{b^2}\frac{{\rm d}^2}{{\rm d}z^2}V^{(\pm)}_{\Delta_{2,1}}(z)
\,+:{\rm T}V^{(\pm)}_{\Delta_{2,1}}(z):\;=\;0\,,
\end{equation}
where
\begin{equation}
:{\rm T}V_{\Delta}(z):
\;\equiv\;\sum\limits_{n\leqslant -2}z^{-n-2}L_{n}V_{\Delta}(z) \,+
\sum\limits_{n\geqslant -1}V_{\Delta}(z)L_{n}z^{-n-2}\;;
\end{equation}

\smallskip\noindent
2.~conformal blocks with $V^{(\pm)}_{\Delta_{2,1}}(z)$
obey certain partial differential equations (PDE's), in particular, 
the 5-point degenerate conformal blocks:\footnote{Here 
$|\,\alpha_i\,\rangle\equiv|\,\Delta_{\alpha_i}\,\rangle$
and the operator--state correspondence is assumed.}
\begin{eqnarray*}
\mathscr{F}_{\pm}(z,x)&\equiv&\langle\alpha_4|V_{\alpha_4,\beta_{\pm}}^{\;\alpha_3}(1)
V^{-b/2}_{\beta_{\pm},\beta}(z)
V_{\beta,\alpha_1}^{\;\alpha_2}(x)|\alpha_1\rangle
\\[5pt]
&=&
\begin{picture}(65,30)(0,5)
\put(0,0){\makebox(15,10)
{$
  \alpha_4
$}}
\put(15,5){\line(1,0){10}}
\put(25,0){\makebox(15,10){$
\scriptstyle 1
$}}
\put(40,5){\line(1,0){40}}
\put(80,0){\makebox(15,10)[c]
{$
 \scriptstyle z
$}}
\put(32.5,10){\line(0,1){8}}
\put(25,20){\makebox(15,10){$
  \alpha_3
$}}
\put(55,8){\makebox(15,10){$
 \scriptstyle \beta_{\pm}
$}}
\put(87,10){\line(0,1){8}}
\put(78,23){\makebox(15,10){$
-\frac{b}{2}
$}}
\put(95,5){\line(1,0){40}}
\put(135,0){\makebox(15,10)[c]
{$
 \scriptstyle x
$}}
\put(108,8){\makebox(15,10){$
 \scriptstyle \beta
$}}
\put(142,10){\line(0,1){8}}
\put(135,23){\makebox(15,10){$
\alpha_2
$}}
\put(147,5){\line(1,0){10}}
\put(158,0){\makebox(15,10){$
 \alpha_1
$}}
\end{picture}
\end{eqnarray*}
obey the following PDE:\footnote{Eqs.~(\ref{QuantumFuchsBlock}) 
are nothing but eq.~(\ref{Fuchs1}) rewritten for conformal blocks.}
\begin{eqnarray}
\label{QuantumFuchsBlock} 
&&\left[\frac{\partial^2}{\partial z^2}
-b^2\left(\frac{1}{z} - \frac{1}{1-z}\right)\frac{\partial}{\partial z}\right.\nonumber
\\
&&\left.
\hspace{50pt}+b^2\left(\frac{\Delta_1}{z^2} +
\frac{\Delta_2}{(z-x)^2}+\frac{\Delta_3}{(1-z)^2} +
\frac{\Delta_1 \!+\! \Delta_2 \!+\!\Delta_3
\!+\!\Delta_{-\frac{b}{2}}\!-\!\Delta_4}{z(1-z)}\right)\right.\nonumber
\\
&&\left.\hspace{100pt}
+b^2\frac{x(1-x)}{z(z-x)(1-z)}
\frac{\partial}{\partial x}\right]\mathscr{F}_{\pm}(z,x)=0.
\end{eqnarray}

In what follows we will compute the classical limit
of eqs.~(\ref{QuantumFuchsBlock}). Let us assume that all `momenta' $\alpha_i$, $\beta$ are heavy, 
i.e.,:\footnote{Here $\beta$ is the same as $\alpha$ in the previous subsection 
and $2\eta=1-\lambda$.}
$\alpha_i=\frac{\eta_i}{b}$ and $\beta=\frac{\eta}{b}$ 
then, in the limit $b\to 0$ one gets
\begin{eqnarray}\label{Factor}
\mathscr{F}_{\pm}(z,x)
%\;\equiv\;\langle\alpha_4|V_{\alpha_4,\beta_{\pm}}^{\alpha_3}(1)
%V^{-b/2}_{\beta_{\pm},\beta}(z)
%V_{\beta,\alpha_1}^{\alpha_2}(x)|\alpha_1\rangle
\;\stackrel{b\to 0}{\sim}\;\Psi_{\pm}\left(\infty,1,z,x,0\right)\;{\rm e}^{\frac{1}{b^2}
f_{\delta}\left[^{\delta_3\,\delta_2}_{\delta_4\,\delta_1}\right](x)},
\end{eqnarray}
where $f_{\delta}\!\left[^{\delta_3\,\delta_2}_{\delta_4\,\delta_1}\right]\!(x)$ is the classical 
spherical 4-point block (\ref{class4pt}) with 
$\delta\!=\!\lim_{b\to 0}b^2\Delta_\beta\!=\!\eta\left(1-\eta\right)$,
$\delta_i\!=\!\lim_{b\to 0}b^2\Delta_i=\eta_{i}(1-\eta_i)$
and
\begin{eqnarray}\label{Psin}
\Psi_{\pm}\left(\infty,1,z,x,0\right) &=& \lim\limits_{b\to 0}
\frac{\langle\alpha_4|V_{\alpha_4,\beta_{\pm}}^{\alpha_3}(1)
V^{-b/2}_{\beta_{\pm},\beta}(z)
V_{\beta,\alpha_1}^{\alpha_2}(x)|\alpha_1\rangle}
{\langle\alpha_4|V_{\alpha_4,\beta}^{\alpha_3}(1)
V_{\beta,\alpha_1}^{\alpha_2}(x)|\alpha_1\rangle}.
%\\
%f_{\delta}\!\left[^{\delta_3\,\delta_2}_{\delta_4\,\delta_1}\right]\!(x)
%&=& \lim\limits_{b\to 0}b^2\log\langle\alpha_4|V_{\alpha_4,\beta}^{\alpha_3}(1)
%V_{\beta,\alpha_1}^{\alpha_2}(x)|\alpha_1\rangle
\end{eqnarray}

To see that the semi-classical asymptotic of $\mathscr{F}_{\pm}(z,x)$ has the factorized form 
(\ref{Factor}) let us define  
for $Z=(z_4, z_3, z)$ the ratio
\begin{equation}\label{ra}
\Psi_{\pm,n}(Z)\equiv\frac{\langle 0|V_{0,\alpha_4}^{\alpha_4}(z_4)V_{\alpha_4,\beta_\pm}^{\alpha_3}(z_3)
V^{-b/2}_{\beta_{\pm},\beta}(z)|\beta^{n}_{I}\rangle}
{\langle 0| V_{0,\alpha_4}^{\alpha_4}(z_4)V_{\alpha_4,\beta}^{\alpha_3}(z_3)|\beta^{n}_{I}\rangle},
\end{equation}
where $|\,\beta_{I}^{n}\,\rangle\equiv|\,\Delta_{\beta,I}^{n}\,\rangle$ 
are the basis vectors of the form (\ref{basis}). Numerical calculations tell us that 
$\Psi_{\pm,n}(Z)$ is light in the limit $b\to 0$, i.e., $\Psi_{\pm,n}(Z)\sim{\cal O}(b^0)$.
Let us take the latter as an assumption. Let 
${\sf X}:=V_{0,\alpha_4}^{\alpha_4}(z_4)V_{\alpha_4,\beta_\pm}^{\alpha_3}(z_3)$, then, 
using (\ref{cr}) and (\ref{ra})
one can compute (cf.~appendix C.2 in \cite{Fitzpatrick:2014vua})
\begin{eqnarray*}
\langle 0|{\sf X} V^{-b/2}_{\beta_{\pm},\beta}(z)L_{-m}|\beta^{n}_{I}\rangle 
&=&\sum_{i=4,3,z}\left[\frac{(m-1)\Delta_i}{z_i}-\frac{1}{z_{i}^{m-1}}\partial_i\right]
\langle 0|{\sf X} V^{-b/2}_{\beta_{\pm},\beta}(z)|\beta^{n}_{I}\rangle
\\[3pt]
&=&\sum_{i=4,3,z}\left[\frac{(m-1)\Delta_i}{z_i}-\frac{1}{z_{i}^{m-1}}\partial_i\right]
\Psi_{\pm,n}(Z)\,\langle 0|{\sf X}|\beta^{n}_{I}\rangle
\\[3pt]
&=&
\Psi_{\pm,n}(Z)\,\langle 0|{\sf X}L_{-m}|\beta^{n}_{I}\rangle
+z^{-1}(m-1)\Delta_{-{b\over 2}}\Psi_{\pm,n}(Z)\,\langle 0|{\sf X}|\beta^{n}_{I}\rangle
\\[3pt]
&-&\sum_{i=4,3,z}\left(z_{i}^{1-m}\partial_i\Psi_{\pm,n}(Z)\right)\langle 0|{\sf X}|\beta^{n}_{I}\rangle\,,
\end{eqnarray*}
where $\partial_i:=\partial_{z_i}$, $z_z:=z$, $\Delta_z:=\Delta_{-{b\over 2}}$.
Dividing both sides of this equation by $\langle 0|{\sf X}L_{-m}|\beta^{n}_{I}\rangle$
we see that the `shifted' ratio 
$$
\Psi_{\pm,n+m}(Z):=\frac{\langle 0|{\sf X} V^{-b/2}_{\beta_{\pm},\beta}(z)L_{-m}|\beta^{n}_{I}\rangle}{
\langle 0|{\sf X}L_{-m}|\beta^{n}_{I}\rangle}=\Psi_{\pm,n}(Z)+\ldots
$$
is also light. Indeed, terms denoted by $\ldots$ are proportional to $\Delta_{-{b\over 2}}$
and derivatives $\partial_i\Psi_{\pm,n}(Z)$. So one may say that the `lightness property'
of the ratios of the type (\ref{ra}) does not depend on the level $n$.
Using this fact and taking into account that 
the generic four-point block with
heavy weights exponentiates in the classical limit,
$$ 
{\cal F}_{c,\Delta_\beta}\!\left[^{\Delta_3\,\Delta_2}_{\Delta_4\,\Delta_1}\right]\!(x)
\equiv\langle\alpha_4|V_{\alpha_4,\beta}^{\alpha_3}(1)
V_{\beta,\alpha_1}^{\alpha_2}(x)|\alpha_1\rangle
\;\stackrel{b\to 0}{\sim}\;
\exp\left\lbrace\frac{1}{b^2}
f_{\delta}\!\left[^{\delta_3\,\delta_2}_{\delta_4\,\delta_1}\right]\!(x)\right\rbrace,
$$
one can compute
\begin{eqnarray*}
\mathscr{F}_{\pm}(z,x)&=&
\langle\alpha_4|V_{\alpha_4,\beta_{\pm}}^{\alpha_3}(1)
V^{-b/2}_{\beta_{\pm},\beta}(z)\mathbb{P}_{\beta}
V_{\beta,\alpha_1}^{\alpha_2}(x)|\alpha_1\rangle
\\[3pt]
&=&
\sum\limits_{n\geq 0}\sum\limits_{I,J\vdash n}\left(G_{\beta}^{(n)}\right)^{IJ}
\langle \alpha_4|V_{\alpha_4,\beta_{\pm}}^{\alpha_3}(1)
V^{-b/2}_{\beta_{\pm},\beta}(z)|\beta_{I}^{n}\rangle
\langle\beta_{J}^{n}|V_{\beta,\alpha_1}^{\alpha_2}(x)|\alpha_1\rangle
\\[3pt]
&=&
\sum\limits_{n\geq 0}\sum\limits_{I,J\vdash n}\left(G_{\beta}^{(n)}\right)^{IJ}
\Psi_{\pm,n}(Z)\,
\langle\alpha_4|V_{\alpha_4,\beta}^{\alpha_3}(1)|\beta_{I}^{n}\rangle
\langle\beta_{J}^{n}|V_{\beta,\alpha_1}^{\alpha_2}(x)|\alpha_1\rangle
\\[3pt]
&\stackrel{b\to 0}{\sim}&
\Psi_{\pm}\left(\infty,1,z,x,0\right)\,
\exp\left\lbrace\frac{1}{b^2}
f_{\delta}\!\left[^{\delta_3\,\delta_2}_{\delta_4\,\delta_1}\right]\!(x)\right\rbrace.
\end{eqnarray*} 
In eqs.~above $(G_{\beta}^{(n)})^{IJ}$ is the inverse of the Gram matrix 
$(G_{\beta}^{(n)})_{IJ}=\langle\,\Delta(\beta)_{I}^{n}\,|\,\Delta(\beta)_{J}^{n}\,\rangle$.\footnote{Here 
and below we use the equivalent notation $\Delta_\beta\equiv\Delta(\beta)$.}
The above calculation defines the functions $\Psi_{\pm}(Z)$, so one can write
\begin{equation}\label{Solutions}
\Psi_{\pm}(\infty,1,z,x,0)=\lim\limits_{b\to 0}\frac{\mathscr{F}_{\pm}(z,x)}
{{\cal F}_{c,\Delta_\beta}\!\left[^{\Delta_3\,\Delta_2}_{\Delta_4\,\Delta_1}\right]\!(x)}\;\;.
\end{equation}

Turning to the problem of the classical limit of eqs.~(\ref{QuantumFuchsBlock}) it is easy to see that
the limit $b\to 0$ taken 
from eqs.~(\ref{QuantumFuchsBlock}) after the substitution  (\ref{Factor}) yields 
the normal form of the Heun ordinary differential 
equation (\ref{Fuchss2bis}) with a pair of Floquet type linearly independent solutions given by 
eq.~(\ref{Solutions}).\footnote{For consistency of our calculations one can check that 
$\lim_{b\to 0}b^2\partial_z\Psi_{\pm}=0$
and $\lim_{b\to 0}b^2\partial_x\Psi_{\pm}=0$.}

In the next section we will look once again into depths of eq.~(\ref{Factor}) in order to
explicitly compute the limit (\ref{Solutions}).

\section{Heavy--light factorization and Floquet type Heun's solutions}
\subsection{Path--multiplicative solutions from conformal blocks}
\label{MR}
In this subsection we analyze the way the degenerate 
five-point blocks factorize in the classical limit. Without loss of generality, let us focus on 
$\mathscr{F}_{+}$. The analysis of the case $\mathscr{F}_{-}$ is analogous. 
Its expansion in terms of intermediate states reads
\begin{multline}
\label{5pt_corr_funct}
\mathscr{F}_{+}(z,x) = 
\bra{\dl_{4}}V_{\dl_{3}}(1)\mathbb{P}_{\dl_{\b_{+}}} V_{\dl_{2,1}}(z)
\mathbb{P}_{\dl_{\b}}  V_{\dl_{2}}(x)\ket{\dl_{1}}
\\
= x^{\dl_{\b}-\dl_{2}- \dl_{1}} z^{\dl_{\b_{+}}-\dl_{2,1}-\dl_{\b}}
\sum_{m,n\geq 0} x^{n}z^{m-n}
\underbrace{\bra{m;\dl_{\b_{+}}} V_{\dl_{2,1}}(1)\ket{n;\dl_{\b}}}_{=:\, 
A_{m,n}}\,,
\end{multline}
where we used the following notation
\begin{subequations}
\begin{equation}
\label{ket_vertex}
\begin{gathered}
V_{\dl_{a}}(x)\ket{\dl_{b}} 
= x^{\dl-\dl_{a}-\dl_{b}} 
\sum_{n\geq 0} x^{n} \ket{n;\dl}\,,
\\
\ket{n;\dl} := \sum_{I\vdash n} \bform{\dl}{\dl_{a}}{\dl_{b}}{I}
L_{-I}\ket{\dl},
\quad
\bform{\dl}{\dl_{a}}{\dl_{b}}{I}
:=\sum_{J\vdash n}\igram{n}{\dl}{I}{J} \vertex{\dl}{\dl_{a}}{\dl_{b}}{J}\,,
\end{gathered}
\end{equation}
and where $\vertex{\dl_{\text{int} }}{\dl_{\text{vtx} }}{\dl_{\text{ext} 
}}{N}
:= \bra{\dl_{\text{int}}}L_{N} 
V_{\dl_{\text{vtx}}}(z)\ket{\dl_{\text{ext}}}\big|_{z\to1}$
takes the following explicit form
\begin{equation}
\label{gamma_vertex}
\vertex{\dl_{\text{int} }}{\dl_{\text{vtx} }}{\dl_{\text{ext} 
}}{N}
= \prod_{i=1}^{\ell(N)}\cbr{\dl_{\text{int} } + k_{i}(N)\dl_{\text{vtx} } - 
\dl_{\text{ext} } + \sum_{s>i}^{\ell(N)}k_{s}(N)}\, .
\end{equation} 
\end{subequations}
Formulas for bra vectors can be inferred from those above. 
The formula \eqref{5pt_corr_funct} can be given yet another form
that is more useful for further study, namely
\begin{subequations}
\begin{equation}
\label{5pt_corr_fun_expandend}
\mathscr{F}_{+}(z,x) = z^{\dl_{\b_{+}}-\dl_{2,1}-\dl_{\b}}
\fourptblock{\dl_{\b}}{\dl_{1}}{\dl_{2}}{\dl_{3}}{\dl_{4}}(x)
\sum_{m\in\mathbb{Z}} z^{m} \chi_{m}(x),
\end{equation} 
where
\begin{equation}
\label{def_chi}
\begin{gathered}
\chi_{m>0}(x) := 
\frac{\sum\limits_{k\geq 0} x^{k} A_{m+k,k}}
{\sum\limits_{k\geq 0} x^{k} \mathcal{F}^{(k)}}\,,
\qquad
\chi_{m<0}(x) := \frac{\sum\limits_{k\geq 0} x^{k-m} 
A_{k-m,k}}{\sum\limits_{k\geq 0} x^{k} \mathcal{F}^{(k)}}\,,
\\
\chi_{m=0}(x) := \frac{\sum\limits_{k\geq 0} x^{k} 
A_{k,k}}{\sum\limits_{k\geq 0} x^{k} \mathcal{F}^{(k)}}\,.
\end{gathered}
\end{equation}
\end{subequations}
The prefactor of eq. \eqref{5pt_corr_fun_expandend} is 
the four-point conformal block defined as
\begin{subequations}
\begin{equation}
\label{4pt_conf_block}
\fourptblock{\dl_{\b}}{\dl_{1}}{\dl_{2}}{\dl_{3}}{\dl_{4}}(x)
=x^{\dl_{\b}-\dl_{2}-\dl_{1}}
\sum_{n\geq 0} x^{n}
\mathcal{F}^{(n)}\,,
\end{equation} 
and
\begin{equation}
\label{4pt_conf_block_coeff}
\mathcal{F}^{(n)}:=\fourptblcoeff{n}{\dl_{\b}}{\dl_{1}}{\dl_{2}}{\dl_{3}}{\dl_{4
} }
:= \sum_{I\vdash n} 
\vertex{\dl_{\b}}{\dl_{3}}{\dl_{4}}{I}
\bform{\dl_{\b}}{\dl_{2}}{\dl_{1}}{I}\,.
\end{equation} 
\end{subequations}
Let us consider for definiteness $\chi_{m>0}$. The matrix element 
$A_{m+n,n}$ has the following structure
\begin{equation}
\label{coeff_Hmgreater0}
A_{m+n,n}
= \sum_{M\vdash n} 
\bra{n+m;\dl_{\b_{+}}}V_{\dl_{{2,1}}}(z)L_{-M}\ket{\dl_{\b}}
\bform{\dl_{\b}}{\dl_{2}}{\dl_{1}}{M}\Big|_{z\to1}\,.
\end{equation}
Developing the matrix element under the sum yields
\begin{multline}
\label{matrix_element}
\bra{m+n;\dl_{\b_{+}}}V_{\dl_{2,1}}(z)L_{-M}\ket{\dl_{\b}}
\\
= \sum_{s=0}^{\ell(M)}\sum_{1\leq i_{1}<\ldots<i_{s}\leq \ell(M)}
(-1)^{s}\bra{m+n;\dl_{\b_{+}}}L_{-M^{\complement}_{s}}
\mbox{ad}_{-M_{s}}V_{\dl_{2,1}}(z)
\ket{\dl_{\b} }
\,,
\end{multline}
where $M^{\complement}_{s}$ is a complement of a partition $M_{s}$ such 
that $M^{\complement}_{s}\cup M_{s}= M$ for any $s$. Explicitly, for 
$1\leq i_{1}<\ldots<i_{s}\leq\ell(M)$ they read
$$
M^{\complement}_{s} = 
(k_{1},\ldots,k_{i_{1}-1},k_{i_{1}+1},\ldots,k_{i_{s}-1},k_{i_{s}+1},\ldots 
k_{\ell(M)})\,,
\quad
M_{s} = (k_{i_{1}},\ldots,k_{i_{s}})\,.
$$
In particular $M^{\complement}_{0} = M_{\ell(M)} = M$, and 
$M_{0} = M^{\complement}_{\ell(M)} = \{0\}$. 
We also denoted
$$
\mbox{ad}_{-M_{s}}V_{\dl_{2,1}}(z)
:= \sbr{L_{-k_{i_{s}}},\ldots\sbr{L_{-k_{i_{1}}},V_{\dl_{2,1}}(z) } }.
$$
In order to get further insight in to the formula \eqref{matrix_element}
we notice that operating the Virasoro algebra element $L_{-m}$ on the state 
$\bra{\dl;n}$ yields
$$
\bra{\dl;n}L_{-m} = (\dl +m\dl_{b}-\dl_{a} + n - m)\bra{\dl;n-m}\,,
\quad
m\leq n.
$$
The generalization of this formula to $L_{-M}:=L_{-k_{1}}\cdots 
L_{-k_{\ell(M)}}$, $|M|=m\leq n$ reads
\begin{equation}
\label{L_on_bra}
\bra{\dl;n}L_{-M} = \vertex{\dl+n-m}{\dl_{b}}{\dl_{a}}{M}\bra{\dl;n-m}\,.
\end{equation} 
Using the above result to the matrix element in eq.  
\eqref{matrix_element} and noticing that $|M^{\complement}_{s}|+|M_{s}| = n$
we obtain the general form of the term that contributes to the expansion 
\eqref{matrix_element}
\begin{multline}
\label{matrix_elem_contrib}
(-1)^{s}\bra{m+n;\dl_{\b_{+}}}L_{-M^{\complement}_{s}}
\mbox{ad}_{- M_{s}}V_{\dl_{2,1}}(z)
\ket{\dl_{\b} }\big|_{z\to 1}
\\
=\fourptblcoeff{m
+|M_{s}|}{\dl_{\b_{+}}}{\dl_{\b}}{\dl_{2,1}}{\dl_{3}}{\dl_{ 4}}\,
\vertex{\dl_{\b_{+}}+m+|M_{s}|}{\dl_{3}}{\dl_{4}}{M^{\complement}_{s}}
\vertex{\dl_{\b}}{\dl_{2,1}}{\dl_{\b_{+}}+m+|M_{s}|}{M_{s}}\,.
\end{multline}
Observe that the first factor in the above equation is $m +|M_{s}|$-th 
coefficient of the  four-point conformal block with a degenerate weight 
$\dl_{2,1}$. The latter is well known to be related to hypergeometric function 
$\hyper$, namely
\begin{equation*}
\label{degenerated_block}
\begin{aligned}
\fourptblock{\dl_{\b_{+}}}{\dl_{\b}}{\dl_{2,1}}{\dl_{3}}{\dl_{4}}(z)
=&z^{\dl_{\b_{+}}-\dl_{2,1}-\dl_{\b}}\sum_{n \geq 0} 
z^{n}
\fourptblcoeff{n}{\dl_{\b_{+}}}{\dl_{\b}}{\dl_{2,1}}{\dl_{3}}{\dl_{4}}
\\
=&
z^{\dl_{\b_{+}}-\dl_{2,1}-\dl_{\b}}
(1-z)^{\dl(\a_{3}-\frac{b}{2})-\dl_{3}-\dl_{2,1} }
\\
&\times
\hyper\cbr{b(\a_{3}-\a_{4}+\bar{\b}-\tfrac{b}{2}),
b(\a_{3}+\a_{4}-\b-\tfrac{b}{2});b(2\bar{\b}-b);z}\,,
\end{aligned}
\end{equation*} 
where $\bar{\a}_{i}=Q-\a_{i}$.
Thus, the coefficient of the conformal block can be read off from the
above relation, which yields ($\dl(\a-\frac{b}{2})-\dl_{\a}-\dl_{2,1} = 
b\a$)
\begin{multline}
\label{Hipergeometric_coeff}
\fourptblcoeff{m+|M_{s}|}{\dl_{\b_{+}}}{\dl_{\b}}{\dl_{2,1}}{\dl_{3}}{\dl_{4}}
= \sum_{k=0}^{m+|M_{s}|}(-1)^{m+|M_{s}|-k}
\\
\times
\frac{(b\a_{3})_{m+|M_{s}|-k}\cbr{b(\a_{3}-\a_{4}+\bar{\b}-\tfrac{b}{2})}_{k} 
\cbr{b(\a_{3}+\a_{4}-\b-\tfrac{b}{2})}_{k}}
{(m+|M_{s}|-k)!k!\cbr{b(2\bar{\b}-b)}_{k} }\,.
\end{multline} 
Combining eqs. \eqref{matrix_elem_contrib}, \eqref{matrix_element} and
\eqref{coeff_Hmgreater0} we arrive at the 
explicit form of coefficient $A_{m+n,n}$, and hence, the nominator of 
$\chi_{m}(x)$ that, when written in a suggestive form, reads
\begin{multline*}
\sum_{n\geq0}x^{n}A_{m+n,n}
=\sum_{n\geq 0}\sum_{M\vdash n} \sum_{s=0}^{\ell(M)}
\sum_{1\leq i_{1}<\ldots<i_{s}\leq \ell(M)}
x^{|M^{\complement}_{s}|}\vertex{\dl_{\b_{+}}+m+|M_{s}|}{\dl_{3}}{\dl_{4}}
{M^{\complement}_{s}}
\bform{\dl_{\b}}{\dl_{2}}{\dl_{1}}{M}
\\
\times x^{|M_{s}|}\fourptblcoeff{m
+|M_{s}|}{\dl_{\b_{+}}}{\dl_{\b}}{\dl_{2,1}}{\dl_{3}}{\dl_{ 4}}\,
\vertex{\dl_{\b}}{\dl_{2,1}}{\dl_{\b_{+}}+m+|M_{s}|}{M_{s}}\,.
\end{multline*}
Instead of summing over levels $n$ we can rearrange 
the sum so that it runs over $|M^{\complement}_{s}|$ and $|M_{s}|$
separately. This rearrangement has an effect in the appearance of the 
symmetry factor multiplying $\bform{\dl_{\b}}{\dl_{2}}{\dl_{1}}{M}$ that comes form 
different ways one subpartition is immersed into 
the other. For instance, keeping summation index $|M^{\complement}_{s}|$ 
and partition $M^{\complement}_{s}$ fixed while summing over $|M_{s}|$ we find that there 
are many terms with the same shape of total partition $M = M^{\complement}_{s} \cup M_{s}$
that differ only by the distribution of parts of the partition $M^{\complement}_{s}$
between parts of $M_{s}$ which occurs provided 
both have equal parts. As a result of this reshuffling of terms in the series 
we get ($|M^{\complement}_{s}|\equiv |I| = r,\,|M_{s}|\equiv |J|=u$)
\begin{multline*}
\sum_{n\geq0}x^{n}A_{m+n,n}
=\sum_{u\geq0}x^{u}\sum_{J\vdash u}\cbr{ \sum_{r\geq0}x^{r}\sum_{I\vdash r}
\pbinom{I\cup J}{I}\,\vertex{\dl_{\b_{+}}+m+u}{\dl_{3}}{\dl_{4}}{I}
\bform{\dl_{\b}}{\dl_{2}}{\dl_{1}}{I\cup J}
}
\\
\times \fourptblcoeff{m+u}{\dl_{\b_{+}}}{\dl_{\b}}{\dl_{2,1}}{\dl_{3}}{\dl_{ 
4}}\,
\vertex{\dl_{\b}}{\dl_{2,1}}{\dl_{\b_{+}}+m+u}{J}
\,,
\end{multline*}
where $\pbinom{I\cup J}{I}$ is the number 
of ways the partition $I$ is immersed
into the partition $I\cup J$. As it may be inferred it equals
\begin{equation}
\label{C_IJ}
\pbinom{I\cup J}{I}:= 
\prod_{i\geq1}\binom{m_{i}(I\cup J)}{m_{i}(I)},
\end{equation} 
where $m_{i}(I\cup J)$ is a multiplicity of a part $i$ in a partition 
$I\cup J$,
whereas $m_{i}(I)$ denotes a multiplicity of $i$ in a partition $I$.
The term in the parentheses can be rewritten in yet another form
\begin{multline}
\label{eq_factorization}
\sum_{n\geq0}x^{n}A_{m+n,n}
=\sum_{u\geq0}x^{u}
\fourptblcoeff{m+u}{\dl_{\b_{+}}}{\dl_{\b}}{\dl_{2,1}}{\dl_{3}}{\dl_{4}}
\\
\times\sum_{J\vdash u}\cbr{ \sum_{r\geq0}x^{r}\sum_{I\vdash r}
\vertex{\dl_{\b_{+}}+m+u}{\dl_{3}}{\dl_{4}}{I}
\bform{\dl_{\b}}{\dl_{2}}{\dl_{1}}{I} 
\pbinom{I\cup J}{I}
\frac{\bform{\dl_{\b}}{\dl_{2}}{\dl_{1}}{I\cup J}}
{\bform{\dl_{\b}}{\dl_{2}}{\dl_{1}}{I}
\bform{\dl_{\b}}{\dl_{2}}{\dl_{1}}{J}}
}
\\
\times\vertex{\dl_{\b}}{\dl_{2,1}}{\dl_{\b_{+}}+m+u}{J}
\bform{\dl_{\b}}{\dl_{2}}{\dl_{1}}{J}\,.
\end{multline}
Observe, that the expression in the parentheses
resembles the contribution to the conformal block (see eq. 
\eqref{4pt_conf_block_coeff}) with the intermediate weight $\dl_{\b}$ shifted
as $\dl_{\b_{+}} = \dl_{\b} + \tfrac{1}{2}+\tfrac{b^{2}}{4} - b\b$, multiplied
by ratio of components of $\b$ forms. In the classical limit 
in which weights scale as
$$
\dl_{i}\sim\d_{i}b^{-2},\ \text{for}\ i= 1,\ldots,4,
\quad \dl_{\b}:=\b\cbr{b+b^{-1}-\b}\sim \d b^{-2},
\quad
\b \sim \et b^{-1},
\quad
c\sim 6 b^{-2}\,,
$$
the term exponentiates to the classical conformal block 
provided the mentioned ratio decouples. 
This occurs if and only if the quotient of components of $\b$ forms
has the following asymptotic behavior
\begin{equation}\boxed{\;
\label{betas_ratio}
\frac{\bform{\dl_{\b}}{\dl_{2}}{\dl_{1}}{I\cup J}}
{\bform{\dl_{\b}}{\dl_{2}}{\dl_{1}}{I}
\bform{\dl_{\b}}{\dl_{2}}{\dl_{1}}{J}}
\overset{b\to0}{\sim}
\frac{1}{\pbinom{I\cup J}{I}}\big(1+\ord{b^{2}}\big)\,.\,}
\end{equation} 
As we have checked in a few cases (cf.~appendix \ref{App_ratio_beta}), this 
indeed takes place.
It means that the expression in parentheses in eq. \eqref{eq_factorization} in the 
classical limit looses the dependence on $J$ and therefore decouples from the series 
over $u$ in the classical limit. The decoupled series exponentiates to the classical 
block, so that we are finally left with the function 
\begin{equation*}
\sum_{n\geq0}x^{n}A_{m+n,n}\overset{b\to0}{\sim}
\ex{\frac{1}{b^{2}}
\clfourptblock{\d}{\d_{1}}{\d_{2}}{\d_{3}}{\d_{4}}(x)}
h_{m}(x)\,,
\end{equation*}
where
$$
h_{m}(x) := \sum_{s\geq0} h_{m,s}(\d_{1},\ldots,\d_{4},\d) x^{s}
=\lim_{b\to0}\chi_{m}(x)\,,
$$
and
$$
h_{m,n}(\d_{1},\ldots,\d_{4},\d)
= \fourptblcoeff{m+n}{\dl_{\b_{+}}}{\dl_{\b}}{\dl_{2,1}}{\dl_{3}}{\dl_{4}}
\fourptblcoeff{n}{\dl_{\b}}{\dl_{1}}{\dl_{2}}{\dl_{2,1}}{\dl_{\b_{+}}+m+n}\Big|_{b=0}
+\ldots
$$
becomes the coefficient of the Heun's function in vicinity of $x=0$.
The Heun's function obtained as a limit $b\to0$ of 
\eqref{5pt_corr_fun_expandend} reads\footnote{Note that 
$1-\eta=\frac{1}{2}(1+\lambda)$ $\Rightarrow$ $\Psi_{+}$
has correct monodromy, cf.~eq.~(\ref{mono}).}
\begin{equation}
\label{def_HeunsFunction}
\Psi_{+}(z,x):=\lim_{b\to 0} 
\frac{\mathscr{F}_{+}(z,x) }
{\fourptblock{\dl_{\b}}{\dl_{1}}{\dl_{2}}{\dl_{3}}{\dl_{4}}(x)}
=z^{1-\eta}\sum_{m\in\mathbb{Z}}z^{m} h_{m}(x)\,,
\end{equation} 
where the coefficients for $m\geq0$ are given in the appendix 
\ref{App_HeunCoeffs}. As for the part of the Heun's function with $m<0$ we get
$$
\sum_{k\geq 0}A_{k-m,k}x^{k-m}
= x^{|m|}\sum_{k\geq 0}A_{k+|m|,k}x^{k},
$$
so that 
\begin{equation}
\label{m_negative}
h_{m<0}(x) = x^{-m}\sum_{s\geq0}h_{-m,s}x^{s}\,,
\end{equation} 
and the coefficients are the same as for $m\geq0$ and are presented in eqs. 
(\ref{h0}--\ref{h3}).

\subsection{Hypergeometric limit}
\label{HLim}
\subsubsection*{Solutions in the limit $x\to 0$}
Let us consider the ration of the matrix element \eqref{5pt_corr_fun_expandend} 
and the four-point conformal block which we 
denote as 
\begin{equation}
\label{H_ratio}
\mathscr{H}_{+}(z,x):=
\frac{\mathscr{F}_{+}(z,x) }
{\fourptblock{\dl_{\b}}{\dl_{1}}{\dl_{2}}{\dl_{3}}{\dl_{4}}(x)}
= z^{b\bar\b} \sum_{m\in\mathbb{Z}}\chi_{m}(x)z^{m}\,,
\end{equation} 
in vicinity of the point $x=0$. As it follows 
form eq. \eqref{def_chi} at this point the only contribution to $\mathscr{H}$ 
comes from $\chi_{m}$ with $m\geq0$. Expansion of this coefficient
about $x=0$ yields
$$
\chi_{m\geq0}(x) 
= \frac{\sum\limits_{k\geq 0} x^{k} A_{m+k,k}}
{1+\sum\limits_{k>0} x^{k} \mathcal{F}^{(n)}}
= A_{m,0}+ \cbr{ A_{m+1,1}- \mathcal{F}^{(1)} A_{m,0} }x
+ \ord{x^{2}}\,.
$$
Hence, at this point it amounts to 
\begin{equation}
\label{positive_m}
\chi_{m>0}(0) = A_{m,0} = 
\bra{m;\dl_{\b_{+}}} V_{\dl_{2,1}}(1)\ket{\dl_{\b}}
= \fourptblcoeff{m}{\dl_{\b_{+}}}{\dl_{\b}}{\dl_{2,1}}{\dl_{3}}{\dl_{4}}\,,
\end{equation} 
so that, taking into account the relationship of the four-point conformal block
with one degenerate weight with the hypergeometric function, $\mathscr{H}_{+}$ from eq. 
\eqref{H_ratio} amounts to 
\begin{align}
\nonumber
\mathscr{H}_{+}(z,0) &= 
\fourptblock{\dl_{\b_{+}}}{\dl_{\b}}{\dl_{2,1}}{\dl_{3}}{\dl_{4}}(z)
\\
&= z^{b\bar\b}(1-z)^{b\a_{3}}
\hyper\bigl(b(\a_{3}-\a_{4}+\bar\b -\tfrac{b}{2})\,,
b(\a_{3}+\a_{4} - \b +\tfrac{b}{2})\,;
b(2\bar\b-b)\,;z
\bigr)\,.
\end{align}
Since at the classical limit parameters $\a_{i}$ 
and $\b$ scale as $\a_{i} \sim \et_{i}/b\,,\b \sim \et/b$, we can immediately 
take the limit $b\to0$ and obtain explicit form of the Heun's 
function at $x=0$ that reads
\begin{equation}
\label{Heun_x0}
\lim_{b\to 0}\mathscr{H}_{+}(z,0)
= z^{1-\eta}(1-z)^{\et_{3}}
\hyper\bigl(\et_{3}-\et_{4} - \et +1\,,
\et_{3}+\et_{4} - \et\,;
2(1-\et)\,;z
\bigr)\,.
\end{equation} 

\subsection*{The hypergeometric equation vs. the Heun equation}
Since the accessory parameter (\ref{c2}) takes the 
explicit form
\begin{equation}\label{c2expan}
c_{2}(x) =  (\d - \d_{1} - \d_{2})\frac{1}{x} + \sum_{n>0} n x^{n-1} 
f^{(n)}\,,
\end{equation}
the content of the square bracket in eq.~(\ref{Fuchss2bis}) amounts in the 
limit $x\to 0$ to 
\begin{eqnarray}
&&
\frac{\delta_1}{z^2}
+\frac{\delta_2}{(z-x)^2} +\frac{\delta_3}{(1-z)^2}+
\frac{\delta_1+\delta_2+\delta_3-\delta_4}{z(1-z)}
+\frac{x(1-x)c_{2}(x)}{z(z-x)(1-z)}\nonumber
\\[3pt]
&&\xrightarrow{x\to0}
\frac{z \delta _3+(z-1) \left(z \delta
   _4-\delta \right)}{(z-1)^2 z^2}\,,
\end{eqnarray}
so that we finally arrive at the following equation
\begin{equation}
\label{HeunEq_x20}
\mathcal{A}\Psi :=
\sbr{
\frac{\dd^{2}}{\dd z^{2}}
+ \frac{z \delta _3+(z-1)z \delta_4 -(z-1)\delta }{(z-1)^2 z^2}
}\Psi =0\,.
\end{equation} 
This equation can be transformed into the form of the hypergeometric equation.
Namely, the transformation 
$U\mathcal{A}U^{-1} U\Psi = \tilde{\mathcal{A}}\tilde{\Psi}$,
where $U = z^{-a}(1-z)^{-b}$ yields
\begin{eqnarray*}
\tilde{\mathcal{A}} &=& \frac{\dd^{2}}{\dd z^{2}} 
+ \frac{2(a - (a+b)z)}{z(1-z)}\frac{\dd}{\dd z}
\\
&+&
\frac{(1-z) z \left[(a+b)\big(1-(a+b)\big) -\delta _4\right]-(1-z) \big(a(1-a)
-\delta \big)-z \big(b(1-b)-\delta
   _3\big)}{(z-1)^2 z^2}\,.
\end{eqnarray*}
The operator $\tilde{\mathcal{A}}$ can be brought to the canonical hypergeometric form
$$
\mathcal{H}:= z(1-z)\frac{\dd^{2}}{\dd z^{2}} + (\g - (\a+\b+1)z)\frac{\dd}{\dd z}
- \a\b\,,
$$
by multiplying by $z(1-z)$, redefining $\d = \et(1-\et),\, \d_{3} = \et_{3}(1-\et_{3})$ 
and setting
\begin{equation}\label{exponentsEq}
a(1-a)-\et(1-\et)=0\,,
\quad
b(1-b)-\et_{3}(1-\et_{3})=0\,.
\end{equation}
The solution of the above equations reads
\begin{equation}
\label{exponents}
a = \et\,\vee\,\ a = 1-\et
\quad
\text{and}
\quad
b= \et_{3}\,\vee\, b = 1-\et_{3}\,
\end{equation} 
such that we obtain
$$
\bar{\mathcal{A}} = z(1-z)\frac{\dd^{2}}{\dd z^{2}} 
+ 2\cbr{a - (a+b)z}\frac{\dd}{\dd z}
+\left((a+b)(1-(a+b))-\delta _4\right)\,.
$$
Depending on the choice of the solution \eqref{exponents} of eqs.~\eqref{exponentsEq}
we get the relationship between the classical block parameters $\et,\,\et_{3},\,\et_{4}$ 
and those from the hypergeomtric operator $\mathcal{H}$, i.e., $\a,\,\b,\,\g$.
For $a = 1-\et$ and $b=\et_{3}$ we have
$$
\a= \eta _3-\eta _4-\eta+1, 
\quad 
\b = \eta _3+\eta _4 -\eta ,
\quad  
\g = 2(1-\et)\,.
$$
Hence, the solution for $a = 1-\et$ and $b=\et_{3}$ canonical at $z=0$ is the hypergeomtric function 
$\hyper$ with the following parameters
$$
\tilde{\Psi}_{+} = \hyper\cbr{\eta _3-\eta _4-\eta+1, \eta _3+\eta _4 -\eta;2(1-\et);z}\,.
$$
Thus, the original function being a solution to eq. \eqref{HeunEq_x20} takes the form
\begin{eqnarray}
\label{Heun_x0_solution}
\Psi_{+} &=& U^{-1}\tilde{\Psi}_{+} 
\\
&=& z^{1-\et}(1-z)^{\et_{3}}\hyper\cbr{\eta _3-\eta 
_4-\eta+1, \eta _3+\eta _4 -\eta;2(1-\et);z}
\quad
\text{for}
\quad
a = 1-\et,\, b=\et_{3}\,.\nonumber
\end{eqnarray} 
The above solution has been labeled with $\Psi_{+}$ to match the one in 
eq.~\eqref{Heun_x0} that stems from the five--point degenerate block in the 
classical limit with a given choice of fusion rule. The second solution of the 
equation \eqref{HeunEq_x20} reads 
\begin{eqnarray*}
\Psi_{-} &=& U^{-1}\tilde{\Psi}_{-} 
\\
&=& z^{\et}(1-z)^{\et_{3}}\hyper\cbr{\eta +\eta _3-\eta _4, \eta +\eta _3+\eta 
_4-1;2\et;z}
\quad
\text{for}
\quad
a = \et,\, b=\et_{3}\,.
\end{eqnarray*}

\subsection{Comparison with other methods} 
The path-multiplicative solutions of the Heun equation can be computed by making use of the 
perturbative methods. 
Such technics have been developed lately by Menotti in \cite{Menotti:2014kra} 
to solve the monodromy problem determining the 
complex-valued accessory parameter in the Heun equation.  A bit modified and generalized but basically the same 
perturbative approach has been used in \cite{HK} to calculate the monodromy representation of the (generalized) 
Heun's {\it opers}.\footnote{In a non-specialist terminology it means nothing else but to compute a fundamental 
system of solutions to the respective differential equations.}

In \cite{Menotti:2014kra} (see also \cite{Menotti:2016jut}) the author study the monodromy problem for the eq.
\begin{equation}\label{M1}
y''(z)+{\cal Q}(z)y(z)=0,
\end{equation} 
where
$$
{\cal Q}(z)=\frac{\delta_1}{z^2}
+\frac{\delta_2}{(z-x)^2} +\frac{\delta_3}{(1-z)^2}+
\frac{\delta_1+\delta_2+\delta_3-\delta_4}{z(1-z)}
+\frac{{\cal C}(x)}{z(z-x)(1-z)}
$$
and $\delta_i=\frac{1}{4}(1-\lambda_{i}^{2})$.
The `potential' ${\cal Q}(z)$ in eq.~(\ref{M1}) is almost the same as in eq.~(\ref{Fuchss2bis}).
The only difference is in the definition of the accessory parameter, namely,
${\cal C}(x)=x(1-x)c_{2}(x)$. Further, it is assumed that ${\cal C}(0)=\delta-\delta_1-\delta_2$.
This assumption is precisely met by $x(1-x)c_{2}(x)$, according to the formula (\ref{c2expan}).
The monodromy problem considered in \cite{Menotti:2014kra}
is the same as stated in the last paragraph of subsection 
\ref{LFT5pt}, i.e., the accessory parameter ${\cal C}(x)$ 
must be adjust so that the monodromy of the fundumantal solutions 
along a contour encircling both $0$ and $x$ has the trace $-2\cos(\pi\lambda)$.
 
To reconstruct the accessory parameter 
${\cal C}(x)$ as a power series in $x$, a specific contour including both the origin $0$ and $x$
is chosen in \cite{Menotti:2014kra} for the monodromy calculation. 
Then, the appropriate fundamental 
solutions $(y^{(1)}(z), y^{(2)}(z))$ are computed in the form 
$y^{(i)}(z)=y^{(i)}_{0}(z)+y^{(i)}_{1}(z)x+\ldots$
by perturbing the Heun equation in small $x$. More precisely, the method relies on an expanding of 
${\cal Q}(z)$ in small $x$, 
${\cal Q}={\cal Q}_{0}+x{\cal Q}_{1}+x^2{\cal Q}_{2}+{\cal O}(x^3)$, 
where in particular one gets 
\begin{eqnarray*}
{\cal Q}_{0} &=&\frac{\delta}{z^2}+\frac{\delta_3}{(z-1)^2}+\frac{\delta_4-\delta_3-\delta}{z(z-1)}\;,\\
{\cal Q}_{1} &=&\frac{2\delta_2-{\cal C}'(0)}{z^2(z-1)}+\frac{2\delta_2-{\cal C}(0)}{z^3(z-1)}\;,\\
{\cal Q}_{2} &=&-\frac{{\cal C}''(0)}{2z^2(z-1)}+\frac{3\delta_2-{\cal C}'(0)}{z^3(z-1)}
-\frac{3\delta_2+{\cal C}(0)}{z^4(z-1)}\;.
\end{eqnarray*} 
The calculation starts with the solutions $(y^{(1)}_{0}(z), y^{(2)}_{0}(z))$ 
to the zeroth-order equation $y''(z)+{\cal Q}_{0}(z)y(z)=0$ and 
then subsequent corrections in powers of $x$ are computed. The accessory parameter ${\cal C}(x)$
thus obtained exactly matches the formula (\ref{c2expan}).

In order to compare the method quoted above with the calculations made 
in previous subsections let us point out that the zeroth-order equation  
is nothing but the eq.~(\ref{HeunEq_x20}) whose solutions are given in terms of 
the hypergeometric functions. In refs.~\cite{Menotti:2014kra,Menotti:2016jut} 
these zeroth-order solutions 
are built out of the hypergeometric functions canonical at $z=1$. An 
argument that appears in \cite{Menotti:2014kra} for such a choice 
is a reasonable observation that `working near $z=0$ is difficult due to the singular nature 
of the kernel' ${\cal Q}(z)$.
On the other hand, we have obtained that in the limit
$x\to0$ the solutions extracted from the conformal blocks yield the zeroth-order solutions built out of the 
hypergeometric functions canonical at $z=0$. In the present work we leave this discrepancy without an 
explanation. It requires a better understanding of a mechanism of an analytic continuation of the solutions 
parallelly within perturbative and CFT approaches.

\section{Concluding remarks}
In the present paper we have studied the mechanism of the heavy-light factorization which occurs in the 
classical limit of conformal blocks with the heavy and light contributions. 
We have examined the factorization property in the case of the simplest 5-point degenerate spherical conformal 
blocks $\mathscr{F}_{\pm}$.
Our goal was to answer the question 
whether this semi-classical asymptotical behavior of $\mathscr{F}_{\pm}$ determines the linearly 
independent Floquet type solutions of the normal form Heun equation.
Recall that the Heun equation in its normal form can be obtained in the classical limit from the null 
vector decoupling equations obeyed by $\mathscr{F}_{\pm}$. 
Analyzing  the semi-classical factorization  in the case under 
consideration we have identified the mechanism 
responsible for the decoupling of the heavy and light contributions. 
In particular, a crucial observation made in the present work is the limit 
(\ref{betas_ratio}) which we have checked in many cases. This 
leads to an interesting novel way of computing the path-multiplicative Heun's functions 
(see the result written down in appendix \ref{App_HeunCoeffs}). 
Indeed, if the observation (\ref{betas_ratio}) 
is true then an analysis performed in subsection \ref{MR} yields a practical method of 
computation of the Floquet type Heun's solutions which is suitable for numerical calculations.
In subsection \ref{HLim} we have computed the $x\to 0$ limit of the Heun's solution
within CFT framework and have confirmed that this has an expected form in terms of the 
hypergeometric function. A demonstration of a complete compatibility of our approach with
the perturbative methods requires further study of the mechanism of the analytic continuation of the 
solutions.We aim to analyze this problem soon. 

Finally, let us stress that methods developed in this work have immediate applications in black hole physics 
problems listed in the introduction. We plan to examine these contexts very soon.
Moreover, we plan to continue an exploration of the correspondence between the Heun equation, the classical 
limit of the null vector decoupling equation for the 5-point function/blocks and the BC$_1$ Inozemtsev 
integrable model. The latter is nothing but the Schr\"{o}dinger spectral problem for a class of elliptic 
so-called Treibich-Verdier potentials built out of the Weierstrass elliptic $\wp$-function.
This spectral problem is yet one more incarnation of the Heun equation
and on the other hand it can be obtained in the classical limit from the null vector decoupling equation
written in elliptic variables living on the torus (cf.~\cite{AKPT} and refs.~therein). 
It turns out that this identification has some 
unspoken so far intriguing consequences for: (i) the theory of elliptic solitons, i.e., the solutions of the 
Korteweg-de Vries equation given by the above-mentioned Treibich-Verdier potentials; (ii)  the study of modular 
properties of ${\cal N} = 2$ gauge theories; (iii) the correspondence between the sphere and torus correlation 
functions in the Liouville theory as well as between the sphere and torus conformal blocks.

\section{Ratio of beta forms in the classical limit\label{App_ratio_beta}}
In what follows we present the results of computations of asymptotic of the 
ratio of 
coefficients of $\b$ in the limit $b\to0$ given in eq. \eqref{betas_ratio} 
which 
we denote here as
$$
R_{I\cup J}
:=\frac{\bform{\dl_{\b}}{\dl_{2}}{\dl_{1}}{I\cup J}}
{\bform{\dl_{\b}}{\dl_{2}}{\dl_{1}}{I}
\bform{\dl_{\b}}{\dl_{2}}{\dl_{1}}{J}}\,.
$$
\begin{itemize}
\item $|I\cup J|=2$, $|I|=1$, $|J|=1$
\begin{eqnarray*}
R_{(1)\cup (1)} &=& \frac{1}{2}
\\
&+&\frac{-2 \left(3 \delta ^2+(5 \delta -3)
   \delta _2\right) \delta _1+(5 \delta
   -3) \delta _1^2+\left(\delta -\delta
   _2\right) \left(\delta  (\delta
   +3)+(3-5 \delta ) \delta _2\right)}{4 \delta  (4 \delta +3)
   \left(\delta -\delta _1+\delta
   _2\right){}^2}b^{2}
\\
&+&\ord{b^{4}}.
\end{eqnarray*}
The value of $\pbinom{(1)\cup (1)}{ (1) } = 2$.

\item $|I\cup J|=3$, $|I|=2$, $|J|=1$
\begin{itemize}
\item[$\blacktriangleright$] $I=(2)\,,J=(1)$
\begin{eqnarray*}
&&R_{(2)\cup (1)} = 1
\\
&+&\frac{2\left(\delta _1-\delta
   _2\right) \left(-2 \left(5 \delta
   ^2+(7 \delta -6) \delta _2\right)
   \delta _1+(7 \delta -6) \delta
   _1^2+\left(\delta -\delta _2\right)
   \left(3 \delta  (\delta +2)+(6-7
   \delta ) \delta _2\right)\right)}{3
   \delta  (\delta +2) \left(\delta
   -\delta _1+\delta _2\right) \left(-3
   \delta _1^2+2 \left(\delta +3 \delta
   _2\right) \delta _1+\left(\delta
   -\delta _2\right) \left(\delta +3
   \delta _2\right)\right)}b^2
\\
&+&\ord{b^{4}}.
\end{eqnarray*}
The value of $\pbinom{(2)\cup(1) }{ (2) } = 1$.

\item[$\blacktriangleright$] $I=(1,1)\,,J=(1)$
\begin{eqnarray*}
R_{(1,1)\cup(1) } &=&
\frac{1}{3}
\\
&+&\frac{ \left(-2 \left(3 \delta ^2+(5
   \delta -3) \delta _2\right) \delta
   _1+(5 \delta -3) \delta
   _1^2+\left(\delta -\delta _2\right)
   \left(\delta  (\delta +3)+(3-5
   \delta ) \delta _2\right)\right)}{3
   \delta  (4 \delta +3) \left(\delta
   -\delta _1+\delta _2\right){}^2}b^2
\\   
&+&\ord{b^{4}}.
\end{eqnarray*}
The value of $\pbinom{(1,1)\cup(1)}{ (1,1) } = 3$.
\end{itemize}

\item $|I\cup J|=4$, $|I|=3$, $|J|=1$
\begin{itemize}
\item [$\blacktriangleright$] $I=(3)\,,J=(1)$
$$
R_{(3)\cup(1) } = 1+\ord{b^{2}}\,,
\qquad
\pbinom{(3)\cup(1)}{(3)} = 1\,.
$$

\item [$\blacktriangleright$] $I=(2,1)\,,J=(1)$
$$
R_{(2,1)\cup (1)} = \tfrac{1}{2}+\ord{b^{2}}\,,
\qquad
\pbinom{(2,1)\cup (1)}{ (2,1) } = 2\,.
$$
\item [$\blacktriangleright$] $I=(1,1,1)\,,J=(1)$
$$
R_{(1,1,1)\cup (1)} = \tfrac{1}{4}+\ord{b^{2}}\,,
\qquad
\pbinom{(1,1,1)\cup (1)}{ (1,1,1) } = 4\,.
$$
\end{itemize}
\end{itemize}
In the last item for $|I\cup J|=4$, because of their sizable forms, we have 
omitted the terms of order $\ord{b^{2}}$.

\section{Coefficients of the Heun's function}
\label{App_HeunCoeffs}
In this appendix we present the explicit form of numerical derivation
of the coefficients of the Heun's function given as
$$
\Psi_{+}(z,x) = z^{1-\eta}\sum_{m\in\mathbb{Z}} h_{m}(x) z^{m},
\quad
h_{m}(x) = \sum_{s\geq0}h_{m,s}(\d_{1},\ldots,\d_{4},\eta) x^{s}\,,
$$
where particular values read
\begin{subequations}
\begin{align}
\label{h0}
h_{0}(x) =& 1-\tfrac{\left(\delta _1-\delta
   _2+(\eta -1) \eta \right) \left(\eta
    \left((\eta
   -1) \eta-\delta _3+\delta _4 -1\right)+2 \delta _3-2
   \delta _4+1\right)}{4 (\eta -1)^2
   \eta }x+\ldots,
\\\nonumber
h_{1}(x) = & \tfrac{\delta _3-\delta _4+(1-\eta )
   \eta }{2 (\eta -1)}
   \\\nonumber
   &+ \tfrac{\left(\delta _1-\delta _2+(\eta
   -1) \eta \right) }{8 (\eta -1)^2 \eta  (2
   \eta -3)}
   \Big(\left(-2
   \delta _3+2 \delta _4+1\right) \eta
   ^3+\left(8 \delta _3-6 \delta
   _4+4\right) \eta ^2
   \\\nonumber
   &+\left(\delta
   _3^2-2 \left(\delta _4+5\right)
   \delta _3+\left(\delta _4-1\right)
   \left(\delta _4+3\right)\right) \eta
   -3 \left(\delta _3-\delta
   _4\right){}^2
   \\
   \label{h1}
   &+3 \left(\delta
   _3+\delta _4\right)+\eta ^4(\eta-3)\Big)x+\ldots
\\[7pt]\nonumber
h_{2}(x) =& \tfrac{\left(5-2 \delta _3+2 \delta
   _4\right) \eta ^2+\left(6 \delta
   _3-4 \delta _4-2\right) \eta +\delta
   _3^2+\left(\delta_4+2\right)(\delta _4 -2 \delta _3)+\eta ^3(\eta-4)}
   {4\left(2 \eta ^2-5 \eta +3\right)}
   \\\nonumber
   &-\tfrac{\left(\delta _1-\delta _2+(\eta
   -1) \eta \right)}{48 (\eta -2) (\eta -1)^2
   \eta  (2 \eta -3)} \Big(\left(-3
   \delta _3+3 \delta _4+16\right) \eta
   ^5+\left(24 \delta _3-18 \delta
   _4-7\right) \eta ^4
   \\\nonumber
   &+\left(3 \delta
   _3^2-3 \left(2 \delta _4+25\right)
   \delta _3+3 \delta _4 \left(\delta
   _4+11\right)-23\right) \eta
   ^3
   \\\nonumber
   &+\left(-21 \delta _3^2+6 \left(6
   \delta _4+19\right) \delta _3-3
   \delta _4 \left(5 \delta
   _4+4\right)+32\right) \eta
   ^2
   \\\nonumber
   &+\left(-\delta _3^3+\left(3 \delta
   _4+49\right) \delta _3^2-3
   \left(\delta _4 \left(\delta
   _4+22\right)+28\right) \delta
   _3+\left(\delta _4-1\right) \delta
   _4 \left(\delta
   _4+18\right)-12\right) \eta 
   \\
   \label{h3}
   &+12\left(2 \delta _3+\delta _4\right)+2
   \left(\delta _3-\delta _4\right)
   \left(2 \delta _3^2-\left(4 \delta
   _4+17\right) \delta _3+2 \delta
   _4^2+\delta _4\right)+\eta ^7-7 \eta
   ^6\Big)x+\ldots
\end{align}
\end{subequations}

\providecommand{\href}[2]{#2}\begingroup\raggedright\endgroup
\end{document}